\documentclass[aps,prd,twocolumn,floatfix,altaffilletter,superscriptaddress,preprintnumbers, tightenlines,showpacs,showkeys,nofootinbib,notoccite]{revtex4-1}
\usepackage[utf8]{inputenc}
\usepackage[colorlinks=true,citecolor=blue,linkcolor=blue]{hyperref}
\usepackage[normalem]{ulem}
\usepackage{amsmath,amssymb, mathrsfs,esvect}
\usepackage{epsfig}
\usepackage{graphicx}               
\usepackage{url}
\usepackage{color}
\usepackage{slashed}
\usepackage{multirow}
\usepackage{hyperref}
\usepackage{placeins}
\usepackage[dvipsnames]{xcolor}
\usepackage{epstopdf}
\usepackage{soul}
\usepackage{tikz}
\usepackage[capitalise, english]{cleveref}
\usepackage{siunitx}
\usepackage{xspace}
\usepackage{booktabs}
\usetikzlibrary{trees}
\usetikzlibrary{decorations.pathmorphing}
\usetikzlibrary{decorations.markings}

\newcommand\myshade{80}
\colorlet{mylinkcolor}{ForestGreen}
\colorlet{mycitecolor}{Red}
\colorlet{myurlcolor}{violet}

\hypersetup{
	linkcolor  = mylinkcolor!\myshade!black,
	citecolor  = mycitecolor!\myshade!black,
	urlcolor   = myurlcolor!\myshade!black,
	colorlinks = true
}

\definecolor{jblue}{RGB}{20,50,100}
\definecolor{npurple}{RGB} {153, 51, 204}
\definecolor{wred}{RGB}{217,0,56}
\definecolor{white}{RGB}{255,255,255}
\definecolor{forestgreen}{HTML}{228B22}

\allowdisplaybreaks
\usepackage{tikz}
\usepackage{tikz-feynman}
\tikzfeynmanset{compat=1.0.0}

\begin{document}
	
\title{Axions from Betelgeuse: Probing Axion Couplings with $^{23}\rm{Na}$ and $^{55}\rm{Mn}$}
\author{Nicholas A. M. Revel}
\email{24hpyl@queensu.ca}
\affiliation{Department of Physics, Engineering Physics \& Astronomy, Queen's University, Kingston, ON, K7L 3N6, Canada}
\affiliation{Arthur B. McDonald Canadian Astroparticle Physics Research Institute, Kingston, ON, K7L 3N6, Canada}

\author{Baldeep Grewal}
\email{baldeep.grewal@queensu.ca}
\affiliation{Department of Physics, Engineering Physics \& Astronomy, Queen's University, Kingston, ON, K7L 3N6, Canada}
\affiliation{Arthur B. McDonald Canadian Astroparticle Physics Research Institute, Kingston, ON, K7L 3N6, Canada}

\author{Anupam Ray}
\email{anupam.ray@queensu.ca}
\affiliation{Department of Physics, Engineering Physics \& Astronomy, Queen's University, Kingston, ON, K7L 3N6, Canada}
\affiliation{Arthur B. McDonald Canadian Astroparticle Physics Research Institute, Kingston, ON, K7L 3N6, Canada}
\affiliation{Perimeter Institute for Theoretical Physics, Waterloo, ON, N2L 2Y5, Canada}

\author{Aaron C. Vincent}
\email{aaron.vincent@queensu.ca}
\affiliation{Department of Physics, Engineering Physics \& Astronomy, Queen's University, Kingston, ON, K7L 3N6, Canada}
\affiliation{Arthur B. McDonald Canadian Astroparticle Physics Research Institute, Kingston, ON, K7L 3N6, Canada}
	
 \begin{abstract}
We investigate axion-like particle (ALP) production in the interior of Betelgeuse via nuclear de-excitation of thermally populated excited states of $^{23}$Na and $^{55}$Mn. As these ALPs propagate to
Earth, they can produce photons either through their conversion in the Galactic magnetic field or through their decay. Using INTEGRAL-ISGRI and NuSTAR observations, we search for these photon signals,
find no significant excess, and derive 95\% CL exclusion limits on the ALP couplings. For magnetic-field conversion of ALPs, the $^{23}\mathrm{Na}$ ($^{55}\mathrm{Mn}$) line observed with INTEGRAL-ISGRI constrains $|g_{a\gamma}g_{ap}| < 4.1\times10^{-21}$ ($6.9\times10^{-19}$)~$\mathrm{GeV}^{-1}$ for $m_a < 10^{-10}~\mathrm{eV}$. For ALP decays, the $^{23}\mathrm{Na}$ ($^{55}\mathrm{Mn}$) line observed with INTEGRAL-ISGRI (NuSTAR) constrains $|g_{a\gamma}g_{ap}|$ down to $3.0\times10^{-22}$ ($2.9\times10^{-19}$)~$\mathrm{GeV}^{-1}$ at $m_a \simeq 400$ ($83$)~keV, weakening in both cases as $m_a^{-3/2}$ for lower ALP masses, and as $m_a^{-2}$ for $^{23}$Na below $m_a \sim 1$~keV, with sensitivity extending to $m_a \simeq 440.2$ ($124$)~keV. Although our $^{23}\mathrm{Na}$-based limits are the strongest for both channels, their larger transition energy makes them more sensitive to the precise evolutionary stage of Betelgeuse, whereas the $^{55}\mathrm{Mn}$-based limits are considerably more robust against such uncertainties.
\end{abstract}
\maketitle

\section{Introduction}
The Standard Model (SM) of particle physics provides a remarkably successful description of the known elementary particles and their interactions \cite{Gaillard_Grannis_Sciulli}. Nevertheless, it does not explain why CP violation in the strong interaction is so strongly suppressed, a puzzle commonly known as the strong CP problem. The Peccei–Quinn mechanism \cite{Pecci_Quinn} offers an elegant solution by introducing a global symmetry whose spontaneous breaking dynamically relaxes the effective CP-violating parameter, with the associated pseudoscalar particle known as the QCD axion. The axion has also motivated the exploration of a broader class of hypothetical pseudoscalar particles, known as axion-like particles (ALPs), which share some of its properties but are not necessarily associated with a solution to the strong CP problem \cite{Jaeckel:2010ni}. Unlike the QCD axion, the masses and couplings of ALPs are generally not subject to the same theoretical relation, allowing for a much broader range of phenomenology. Depending on their masses and coupling strengths, ALPs can interact with photons, nucleons, and electrons, giving rise to a rich phenomenology that can be explored through complementary laboratory experiments, astrophysical observations, and cosmological probes~\cite{Raffelt:2006cw,Jaeckel:2010ni,Marsh:2015xka,Carenza:2024ehj}. 
Among the astrophysical probes of the ALP–nucleon coupling, an intriguing possibility was proposed by Haxton and Lee \cite{Haxton:1991pu}, who first explored the production of axions through nuclear de-excitation transitions in stellar environments. Thermally populated excited nuclear states can undergo radiative transitions, emitting ALPs through their coupling to nucleons. These ALPs can escape the stellar interior and, while propagating toward Earth, may produce observable photons either through conversion into photons in astrophysical magnetic fields or through their decay into two photons, depending on their mass and coupling. The former process can generate a nearly monochromatic photon signal, with the photon energy determined by the nuclear transition that produced the ALPs, while the latter can yield a characteristic photon spectrum whose properties depend on the ALP mass and decay kinematics \cite{Raffelt:1987im,Raffelt:1996wa}.

Previous searches have primarily considered solar ALP production through nuclear de-excitation, particularly the 14.4~keV transition of thermally excited $^{57}$Fe nuclei in the solar interior~\cite{Moriyama,Krcmar:1998xn,Namba:2007rm,Derbin:2009yb,Andriamonje2009CAST,CUORE:2012ymr,DiLuzio:2021qct,Kumar:2026nlg}. The temperature of the solar core ($\sim$ 1.3 keV) ensures a non-negligible thermal population of the excited nuclear state at 14.4~keV, which, together with the proximity of the Sun to Earth, makes this transition an interesting avenue for probing ALP interactions. Extending these searches beyond the Sun, recent studies have proposed hot, magnetized white dwarfs \cite{Fleury:2022plh}, Betelgeuse \cite{Cand_n_2025}, and nearby galaxies such as M82 and M87 \cite{Ning:2025kyu} as potential sources of mono-energetic ALPs via $^{57}$Fe nuclear transitions. Significant constraints on ALP interactions have been derived by exploiting $a \to \gamma$ conversion in Galactic magnetic fields. 

\begin{figure}[t]
\centering
\includegraphics[width=\columnwidth]{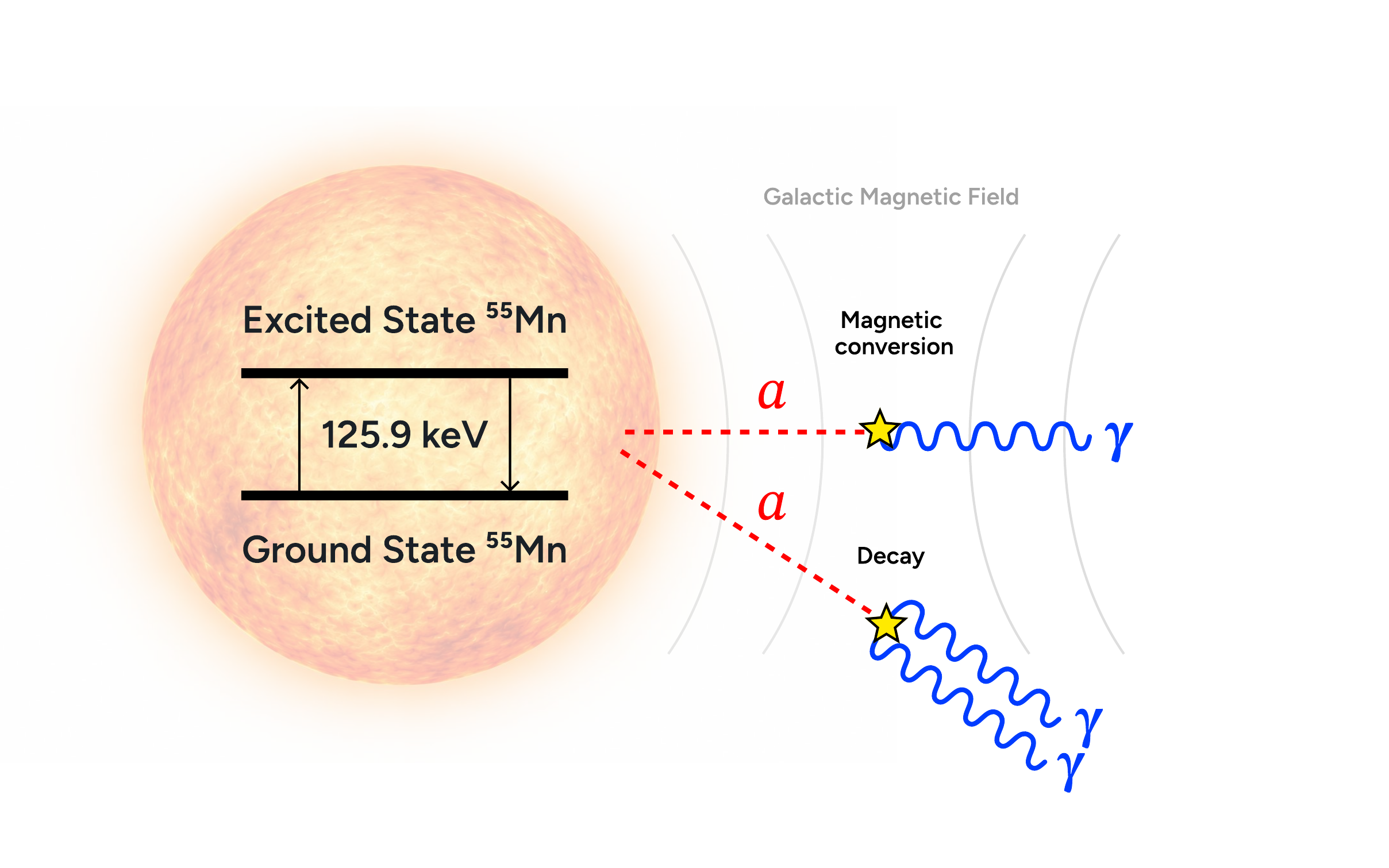}
\caption{Illustration of the ALP being produced from $^{55}\mathrm{Mn}$ de-excitation in Betelgeuse. The produced ALPs can convert into photons in the Galactic magnetic field or decay into two photons, giving rise to observable signals.}
\label{fig:illustration}
  \end{figure}

Several other nuclei that are relatively abundant in stars exhibit nuclear de-excitation transitions at higher energies, making them less accessible for solar ALP searches due to the exponentially suppressed thermal population of their excited states \cite{Ning:2025kyu}.  Hotter stars, however, can provide more favourable conditions for populating these states, enabling the production of higher-energy mono-energetic ALPs and opening up complementary opportunities to probe ALP interactions across a broader energy range. Motivated by this possibility, we investigate ALP production via nuclear de-excitation of two isotopes: $^{55}$Mn and $^{23}$Na~\footnote{Carbon-burning stars in the Milky Way have been proposed as a novel source of ALPs produced via nuclear de-excitation of $^{23}$Na~\cite{Haxton_Liu_McCutcheon_Ray}. See also~\cite{Haxton:2026ura, Haxton:2026pty} for a tantalizing detection possibility of such ALPs with existing NaI detector arrays.}, which are relatively abundant in Betelgeuse and whose excited nuclear states can be thermally populated owing to the star's high temperature. While a recent study \cite{Cand_n_2025} explored ALP production via $^{57}$Fe nuclear transitions in Betelgeuse, our work focuses on $^{55}$Mn and $^{23}$Na, which, unlike $^{57}$Fe are \textit{odd-proton} nuclei. This distinction allows us to primarily probe the ALP–proton coupling, complementing existing constraints on the ALP–neutron coupling from $^{57}$Fe nuclear transitions. Moreover, the higher excitation energies of these two nuclei (125.9 and 440.2~keV, compared with 14.4~keV for $^{57}$Fe) allow ALPs up to these masses to be produced, and since the decay rate grows with the ALP mass, this yields more stringent limits in the decay channel.
We investigate the detectability of these mono-energetic ALPs through their conversion into photons in the Galactic magnetic field, which is efficient for sufficiently small ALP masses. For higher masses, we consider their decays into two photons, which become increasingly relevant owing to the mass dependence of the decay rate. We search for these photon signals in public INTEGRAL-ISGRI and NuSTAR data. We find no significant excess and place upper bounds on the coupling products $|g_{a\gamma}g_{ap}|$ and $|g_{a\gamma}g_{an}|$, where $g_{a\gamma}$, $g_{ap}$ and $g_{an}$ are the ALP couplings to photons, protons and neutrons. A schematic diagram of the processes considered in this work is shown in Fig.~\ref{fig:illustration}.
The rest of the paper is organized as follows. In Sec.~\ref{ALPPRODUCTION}, we estimate the ALP production rate from Betelgeuse via nuclear transitions of $^{55}$Mn and $^{23}$Na. In Sec.~\ref{ALP_CONVERSION}, we briefly review the magnetic field conversion for ultralight ALPs and their decay to two photons for heavier ALPs. In Sec.~\ref{instrumental_analysis}, we discuss the relevant instrumental details and the data analysis procedures. In Sec.~\ref{Results}, we present our results, and in Sec.~\ref{conclusion} we conclude. Since our limits depend on two inputs with considerable
uncertainties, we show their dependence on both in the appendices: in App.~\ref{bfield_uncertainty} we vary the Galactic magnetic field between $B_T=1$ and $3~\mu$G, and in App.~\ref{uncertainty} we compare the limits for Betelgeuse in early core helium burning and at the end of core carbon burning, and show how the ALP emission rate changes across all intermediate evolutionary stages. Throughout, we
set $c=\hbar=k_B=1$.

\section{ALP Production via Nuclear De-excitation}
\label{ALPPRODUCTION}
We consider the ALP–nucleon interaction Lagrangian
\begin{equation}
\mathcal{L}^{\mathrm{int}}_{a} \supset \frac{1}{2m_n} \bar{N}\gamma^\mu\gamma_5 \left(g_{aNN}^{0}+g_{aNN}^{3}\tau_3\right)
N\,\partial_\mu a\,,
\end{equation}
where $\tau_3$ is the third component of isospin, $m_n$ is the nucleon mass, $g_{aNN}^{0} = (g_{ap}+g_{an})/2$ denotes the isoscalar coupling, and $g_{aNN}^{3} = (g_{ap}-g_{an})/2$ denotes the isovector coupling.

\begin{figure}[t]
      \centering
      \includegraphics[width=\columnwidth]{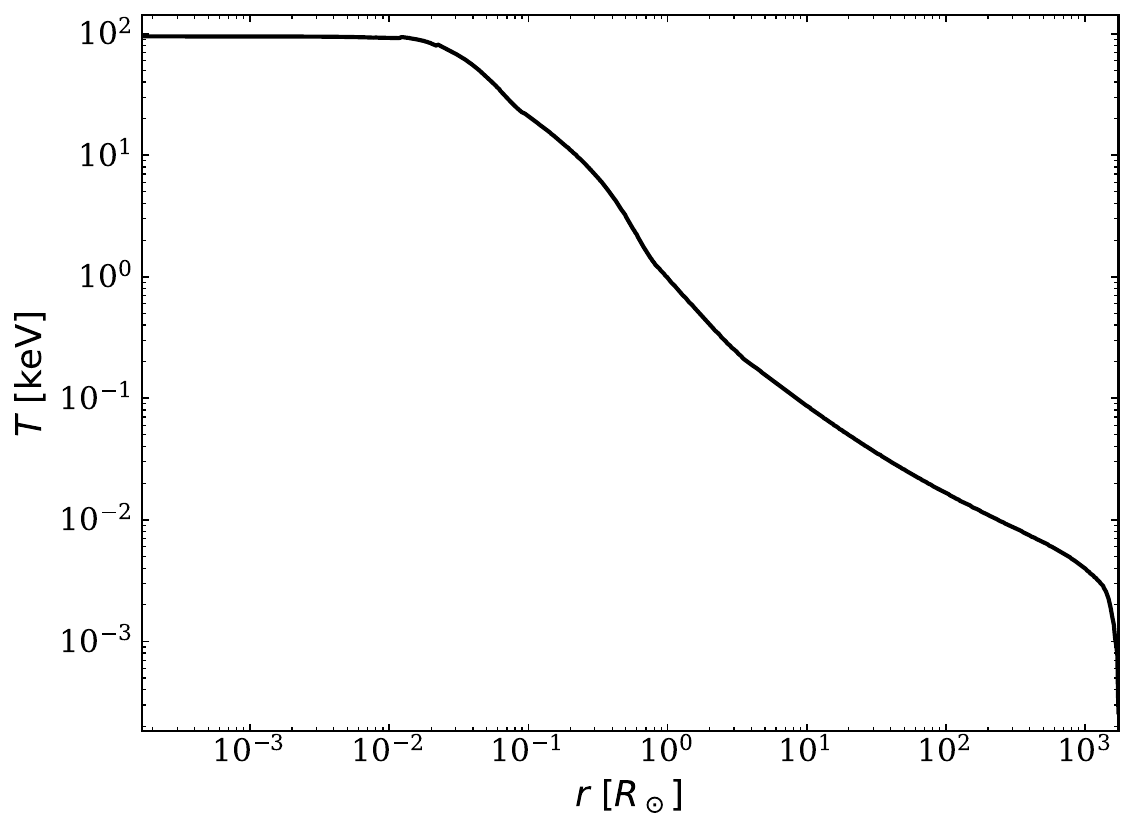}
      \caption{Temperature profile of the Betelgeuse MESA model
      ($M_{\rm init} = 20\,M_\odot$, $Z = 0.02$, end of core carbon burning) as a
      function of the stellar radius.}
      \label{fig:temperature_profile}
  \end{figure}

The ALP–nucleon interaction allows an excited nucleus to emit an ALP as it de-excites. The corresponding ALP emission rate density can be expressed as~\cite{Haxton:1991pu}
 \begin{equation}
      \mathcal{\dot N}_{a}(r) = \mathcal{N}(r)\,\omega_{1}(T(r))\, \frac{1}{\tau_{0}}\, \frac{1}{1+\alpha}\, \frac{\Gamma_{a}}{\Gamma_{\gamma}},
      \label{eq:emission_rate}
  \end{equation}
where $\mathcal{N}(r)$ is the number density of the relevant isotope, $\omega_{1}(T(r))$ is the dimensionless thermal occupation fraction of the nuclear excited state where $T(r)$ is the temperature profile of Betelgeuse (taken from Fig.~\ref{fig:temperature_profile}), $\tau_0$ is the mean lifetime of the excited state, and $\alpha$ is the internal conversion coefficient. The ratio $\Gamma_a/\Gamma_\gamma$ quantifies the relative probability of emitting an ALP rather than a photon in the transition. This ratio depends on the multipolarity of the transition and for magnetic dipole (M1) transitions, it can be expressed as~\cite{Haxton:1991pu}
\begin{equation}
      \label{eq:branching_fraction}
       \frac{\Gamma_{a}}{\Gamma_{\gamma}}
      = \left(\frac{p_a}{p_{\gamma}}\right)^3 \frac{1}{2\pi \alpha_{\rm EM}} \frac{1}{1 + \delta^2}
        \left( \frac{g^0_{aNN} \beta + g^3_{aNN}}{\left( \mu_0 - \frac{1}{2} \right) \beta + \mu_1 - \eta} \right)^2\,.
  \end{equation}
In the above equation, $p_a = \sqrt{E_{*}^2 - m_a^2}$ and $p_\gamma = E_*$ denote the momenta of the outgoing ALP and photon, respectively, where $E_*$ is the nuclear transition energy. $\alpha_{\rm EM} = 1/137$ is the fine-structure constant, $\delta$ is the electric quadrupole (E2) to magnetic dipole (M1) mixing ratio, which can be obtained from~\cite{Nudat3}, and $\mu_0 (\mu_1) = 0.88 (4.706)$ are the dimensionless isoscalar and isovector nucleon magnetic moments~\cite{ParticleDataGroup:2022pth}. The coefficients $\beta$ and $\eta$ are nuclear-structure-dependent parameters calculated within the nuclear shell model~\cite{Haxton:1991pu}. See Table~\ref{tab:parameters_1} for the numerical values of the nuclear-structure parameters and other relevant quantities for the transitions considered in this work.

The thermal occupation fraction of the nuclear excited state at temperature $T(r)$ is given by 
\begin{equation}
      \omega_{1}(T(r)) = \frac{(2J_{1} + 1)e^{-E_{*}/T(r)}}{(2J_{0} + 1) + (2J_{1} + 1)e^{-E_{*}/T(r)}},
      \label{eq:occupation_num}
  \end{equation}
where $E_{*}$ is the nuclear transition energy, and $J_0 (J_1)$ are the angular momenta of the ground (excited) state. 
  
\begin{table}[h]
    \centering
    \begin{tabular}{|c|c|c|c|}
    \hline
    \textbf{Parameter} & \textbf{${}^{55}\mathrm{Mn}$} & \textbf{${}^{23}\mathrm{Na}$} & \textbf{${}^{57}\mathrm{Fe}$} \\
    \hline
    $\beta$      & 0.79   & 0.88   & $-1.19$ \\
    $\eta$       & $-3.74$ & $-1.20$   & 0.80 \\
    $\delta$     & 0.052  & 0.065  & 0.00223 \\
    $J_{0}$      & 5/2    & 3/2    & 1/2 \\
    $J_{1}$      & 7/2    & 5/2   & 3/2 \\
    $\tau_{0}$ [s] & $3.74 \times 10^{-10}$ & $1.64 \times 10^{-12}$ & $1.42 \times 10^{-7}$ \\
    $\alpha$     & 0.0174 & 0 & 8.56 \\
    $E_{*}$ [keV] & 125.9 & 440.2 & 14.4 \\
    \hline
    \end{tabular}
    \caption{The numerical values of relevant parameters required for ALP emission via nuclear de-excitation. The nuclear-structure factors $\beta$ and $\eta$ are taken from~\cite{Haxton:1991pu} and all the other parameters are taken from~\cite{Nudat3}.}
    \label{tab:parameters_1}
    \end{table}
Using the relevant parameters listed in Table~\ref{tab:parameters_1}, we calculate the branching ratios for the isotopes considered in this work (assuming $m_a \ll E_*$). For $^{23}$Na, this evaluates to 
\begin{equation}
\frac{\Gamma_{a}}{\Gamma_{\gamma}} = 0.49(g_{ap}-0.064g_{an})^2 = 0.49 \left (g^{\rm eff,^{23}Na}_{aNN} \right)^2\,.
\end{equation}
For $^{55}$Mn, this is
\begin{equation}
\frac{\Gamma_{a}}{\Gamma_{\gamma}} = 0.23(g_{ap}-0.117g_{an})^2 = 0.23\left (g^{\rm eff,^{55}Mn}_{aNN} \right)^2,
\end{equation}
and for $^{57}$Fe,
\begin{equation}
\frac{\Gamma_{a}}{\Gamma_{\gamma}} = 2.19(g_{an}+0.087g_{ap})^2 = 2.19\left (g^{\rm eff,^{57}Fe}_{aNN} \right)^2\,.
\end{equation}
Clearly, the relative sensitivity to the proton and neutron couplings depends on the nuclear structure of each isotope. For $^{23}$Na and $^{55}$Mn, both odd-proton nuclei, the branching ratios are dominated by the proton coupling $g_{ap}$, whereas for $^{57}$Fe, an odd-neutron nucleus, the neutron coupling $g_{an}$ provides the dominant contribution. The total ALP production rate is then obtained by integrating the local ALP emission rate density over the volume of the star
\begin{equation}
      \frac{\mathrm{d}\dot N_{a}(E_{a})}{\mathrm{d}E_{a}}
      = \int_{0}^{R_{\text{star}}}
      \mathcal{\dot N}_{a}(r)\,\mathcal{G}\bigl(E_{*},\sigma_{\rm eff}[T(r)]\bigr)\,4\pi r^{2}\,\mathrm{d}r,
      \label{eq:production_rate}
  \end{equation}
where $\mathcal{\dot N}_a(r)$ is the local ALP emission rate density defined in Eq.~\eqref{eq:emission_rate} and $R_{\rm{star}}$ is the stellar radius. $\mathcal{G}$ describes the thermally broadened Gaussian energy distribution of the emitted ALPs, centred at the line energy $E_{*}$ with an effective width $\sigma_{\text{eff}}$ determined by the stellar temperature profile $T(r)$.
The largest contribution to the effective width, $\sigma_{\rm eff}$, arises from the thermal broadening of the emitted line, and is given by~\cite{Cand_n_2025}
\begin{equation}
\sigma_{\rm eff} \equiv \sigma_{\text{th}}[T(r)] = E_{*} \sqrt{\frac{ T(r)}{m_{N}}}\,,
    \label{eq:sigma_th}
  \end{equation}
where $m_{N}$ is the mass of the relevant nucleus, and $T(r)$ is the stellar temperature profile shown in Fig.~\ref{fig:temperature_profile}. While stellar rotation further broadens the emitted ALP energy, its contribution of $\sim 0.2$ eV is negligible compared to the thermal broadening~\cite{Cand_n_2025}.

Since $T(r)\leq T_{\rm core}\simeq95$~keV (Fig.~\ref{fig:temperature_profile}), Eq.~(\ref{eq:sigma_th}) yields a thermal width of $\sigma_{\rm th}\lesssim19$~eV for $^{57}$Fe. This is an order of magnitude smaller than the NuSTAR
energy resolution, $\sigma_{\rm det}\simeq0.20$~keV (corresponding to 0.46~keV FWHM) at 14.4~keV, as determined from the response matrix of the corresponding NuSTAR observation~\cite{NuSTAR:2013yza, Madsen:2015jea}. Similarly, for $^{55}$Mn and $^{23}$Na we find $\sigma_{\rm th}\lesssim0.2$ and 0.9~keV, well below the INTEGRAL-ISGRI energy resolution. From the OSA (INTEGRAL Off-line Scientific Analysis) response matrices of the ISGRI pointings used here, a signal at 125.9~keV ($^{55}$Mn de-excitation energy) has a FWHM of $\simeq16$~keV ($\sigma_{\rm det}\simeq6.9$~keV), and a signal at 440.2~keV ($^{23}$Na de-excitation energy) a FWHM of $\simeq39$~keV ($\sigma_{\rm det}\simeq17$~keV). Thus, in all three cases, the thermal broadening is subdominant to the instrumental resolution, making the effective line contribution minimal.

For the stellar simulation used throughout this analysis we use the MESA (Modules for Experiments in Stellar Astrophysics)~\cite{MESA1} evolutionary model of a $20\,M_{\odot}$ star with metallicity $Z=0.02$, meant to simulate Betelgeuse. The simulation was evolved from the pre-main sequence through core hydrogen and helium burning into core carbon burning, following the procedure of
Ref.~\cite{Cand_n_2025}. Unless stated otherwise, we use the profile at the end of core carbon burning; the dependence of the coupling constraints on the evolutionary stage is shown in App.~\ref{uncertainty}.

\section{ALP detection: magnetic field conversion and decay}
\label{ALP_CONVERSION}
After being produced in the stellar interior, the mono-energetic ALPs escape the star and can be probed through two distinct detection channels, depending on their mass. For ultralight ALPs, conversion into photons in the Galactic magnetic field can produce a narrow X-ray spectral line, whereas, for heavier ALPs, their decay into two photons can generate a detectable X-ray signal. See Fig.~\ref{fig:illustration} for a schematic diagram of these two processes. In the following, we briefly review the magnetic field conversion and decay of ALPs.
\label{probability}
\subsection{Magnetic field conversion of ALPs: $a \rightarrow \gamma$} 
Mono-energetic ALPs produced via nuclear de-excitation can convert into photons as they propagate through the Galactic magnetic field. For an ALP of energy $E_a$ traversing a coherent magnetic-field region of length $D = 197$~pc, the ALP-to-photon conversion probability is given by \cite{Raffelt:1987im, Haxton_Liu_McCutcheon_Ray}
\begin{equation}
P_{a\to\gamma} = (\Delta_{a\gamma} D)^2
\frac{\sin^2(\Delta_{\mathrm{osc}} D/2)}
{(\Delta_{\mathrm{osc}} D/2)^2}
\label{eq:conv_prob}
\end{equation}
where
\[
\Delta_{\rm osc}
=
\left[
\left(\Delta_a-\Delta_{\rm pl}\right)^2
+4\Delta_{a\gamma}^2
\right]^{1/2},
\qquad
\Delta_{a\gamma}
=
\frac{g_{a\gamma}B_T}{2},
\]
\[
\Delta_a=-\frac{m_a^2}{2E},
\qquad
\Delta_{\rm pl}=-\frac{\omega_{\rm pl}^2}{2E},
\]

where $B_T$ is the component of the magnetic field transverse to the ALP propagation direction and $\omega_{\mathrm{pl}} = \sqrt{4\pi \alpha_{\rm EM} n_e/m_e}$ denotes the plasma frequency. We model the Galactic medium along the Betelgeuse line of sight as having a spatially uniform transverse magnetic field,
$B_T\simeq1\,\mu{\rm G}$ \cite{Unger:2023lob} with free electron density $n_e\simeq0.011$~cm$^{-3}$ \cite{Haxton_Liu_McCutcheon_Ray}.
Substituting these into
Eq.~\eqref{eq:conv_prob} yields \cite{Haxton_Liu_McCutcheon_Ray}
\begin{align}
\Delta_a &= -7.8 \times 10^{-3}
\times \left(\frac{m_a}{10^{-11}\,\mathrm{eV}}\right)^2
\left(\frac{E}{\mathrm{MeV}}\right)^{-1}
\mathrm{kpc}^{-1},
\\
\Delta_{\mathrm{pl}} &= -1.1 \times 10^{-3}
\times \left(\frac{E}{\mathrm{MeV}}\right)^{-1}
\mathrm{kpc}^{-1},
\\
\Delta_{a\gamma} &= 1.5 \times 10^{-2}
\times \left(\frac{g_{a\gamma}}
{10^{-11}\,\mathrm{GeV}^{-1}}\right)
\left(\frac{B_T}{10^{-6}\,\mathrm{G}}\right)
\mathrm{kpc}^{-1}.
\end{align}
The expected photon flux at Earth is then
  \begin{equation}
   \frac{\mathrm{d}\Phi_\gamma(E_\gamma)}{\mathrm{d}E_\gamma}=\frac{1}{4\pi D^2}\times
   \frac{\mathrm{d}\dot N_a(E_a)}{\mathrm{d}E_a}\bigg|_{E_a=E_\gamma}\times P_{a\to\gamma}(E_\gamma).
  \end{equation}

\subsection{ALP decay: $a\!\to\!\gamma\gamma$ }
 Mono-energetic ALPs produced via nuclear de-excitation can be converted into photons through ALP decay. Since the ALP decay length can be comparable to or larger than the distance $D$ between Earth and Betelgeuse, decay photons can reach Earth from directions away from the star. For an ALP of energy $E_a$, the differential flux of photons at Earth with energy $E_\gamma$ and angle $\beta$
  from Betelgeuse is given as \cite{Candon:2024eah}
 \begin{equation}
  \begin{aligned}
  \frac{\mathrm{d}\Phi_\gamma(E_\gamma)}{\mathrm{d}E_\gamma\,\mathrm{d}\beta}                                                                                           
  &=
  \int_{E_{a,\min}}^{+\infty}
  \frac{2\,\mathrm{d}E_a\cos\beta}{p_a}
  \frac{\mathrm{d}\dot{N}_a/\mathrm{d}E_a}{4\pi D}
  \\
  &\quad\times
  \Theta(\cos\beta-\cos\theta)
  \exp\left[
  -\frac{\Gamma_a m_a D\sin\beta}
  {p_a\sin\theta}
  \right]
  \frac{\Gamma_a m_a}{p_a\sin\theta},
  \end{aligned}
  \label{eq:decay_general}
  \end{equation}
  where $m_{a}$, $p_{a}$ and $v_a=p_a/E_a$ are the ALP mass, momentum and velocity respectively,
  $\Gamma_a = g_{a\gamma}^2 m_a^3/64\pi$ is the rest-frame decay rate, $\Theta$ is the Heaviside step
  function, $E_{a,\min}=E_\gamma+m_a^2/4E_\gamma$ is the minimum ALP energy that can produce a
  photon of energy $E_\gamma$, $D$ is the distance from Betelgeuse to Earth and $\theta$
  is the angle between the ALP and photon directions at decay, with \cite{Candon:2024eah}
  \begin{equation}
  \cos \theta = \frac{1}{v_a}
  \left[
  1 - \frac{E_a}{2E_\gamma}(1 - v_a^2)
  \right].
  \end{equation}
  The ALP spectrum $d\dot N_a/dE_a$ is given by Eq.~\eqref{eq:production_rate}. A photon arriving at angle $\beta$
  originates from an ALP that decayed at a distance $\ell=D\sin\beta/\sin\theta$ from Betelgeuse,
  so that the exponential in Eq.~\eqref{eq:decay_general} is the probability $e^{-\ell/L_a}$ that the ALP survives up to that point,
  with decay length $L_a=p_a/(\Gamma_a m_a)$. We assume that photons from ALPs decaying inside the star ($\ell<R_{\rm star}$) are absorbed, while
  those from decays outside the star reach Earth unattenuated. The signal is collected within an
  angular radius $\beta_{\rm ap}$ of Betelgeuse. For NuSTAR this is the source extraction region,
  $\beta_{\rm ap}=60''$. Our ISGRI spectra are the OSA point-source spectra of Betelgeuse, so only photons within the point-spread function contribute; we take $\beta_{\rm ap}=6'$, half the $12'$ FWHM angular resolution of IBIS. In both cases $\beta_{\rm ap}\ll1$, so we
  can take $\cos\beta\simeq1$ and $\sin\beta\simeq\beta$. The Heaviside function further restricts
  the arrival angle to $\beta<\theta$, so photons emitted close to the ALP direction
  ($\theta<\beta_{\rm ap}$) only reach the detector from within an angle $\theta$ of Betelgeuse.
  Integrating Eq.~\eqref{eq:decay_general} over $R_{\rm star}\sin\theta/D<\beta<\beta_{\max}$, with
  $\beta_{\max}=\min(\beta_{\rm ap},\theta)$, gives the expected photon flux at Earth as
  \begin{equation}
  \frac{d\Phi_\gamma(E_\gamma)}{dE_\gamma}
  =
  \int_{E_{a,\min}}^{+\infty}
  \frac{2\,dE_a}{p_a}
  \frac{d\dot{N}_a/dE_a}{4\pi D^2}\,
  P_{a\to\gamma\gamma},
  \label{eq:decay_flux}
  \end{equation}
  with
  \begin{equation}
  P_{a\to\gamma\gamma}
  =
  \exp\left(-\frac{\Gamma_a m_a R_{\rm star}}{p_a}\right)
  -\exp\left(-\frac{\Gamma_a m_a D\sin\beta_{\max}}{p_a\sin\theta}\right).
  \label{eq:P_decay}
  \end{equation}

\section{Instrumental details and data analysis}
\label{instrumental_analysis}
\subsection{Instrumental details}
\label{ALP_DETECTION}
To search for ALP-induced photon signals from Betelgeuse, we consider two instruments: NuSTAR and ESA’s INTEGRAL-ISGRI. Both instruments are well suited to our search, as together they cover the photon energies relevant to our analysis. In the following, we briefly review the relevant instrumental characteristics, observational details, and data analysis procedure adopted in this work.

\label{data_analysis}
\subsubsection{NuSTAR}
NuSTAR is a focusing X-ray telescope with CdZnTe detectors operating in the 3--79~keV range~\cite{NuSTAR:2013yza}, which makes it well suited to search for the $^{57}$Fe signal, through
both magnetic conversion and decay, and the $^{55}$Mn signal, through decay. It has two co-aligned focal-plane modules, FPMA and FPMB, each with a field of view of $13'\times13'$ and an angular resolution of $58''$ half-power diameter~\cite{NuSTAR:2013yza}. We use the NuSTAR observation of Betelgeuse
taken on August 23, 2019 (ObsID 30501012002). The data are processed with HEASoft (version 6.36) and NuSTARDAS (version 2.1.4)~\cite{2014ascl.soft08004N}. Using the science-mode (mode 01) data, the exposure times are 49.16~ks for FPMA and 48.83~ks for
FPMB. We extract the source spectrum from a circular region of radius $60''$ centred on Betelgeuse, RA $=88^{\circ}47'34.8''$ and Dec $=+07^{\circ}24'25.4''$ (J2000)~\cite{Cand_n_2025}, and the background spectrum from a source-free polygonal region of $\simeq1.8$ times the area, located $5'$--$8'$ from the source on the same detector chip. The expected counts are obtained by forward-folding the ALP-induced photon spectrum through the response files and exposure time of each module~\cite{Cand_n_2025}.

\subsubsection{INTEGRAL}
\begin{figure*}[t] 
    \centering
    \includegraphics[width=\textwidth]{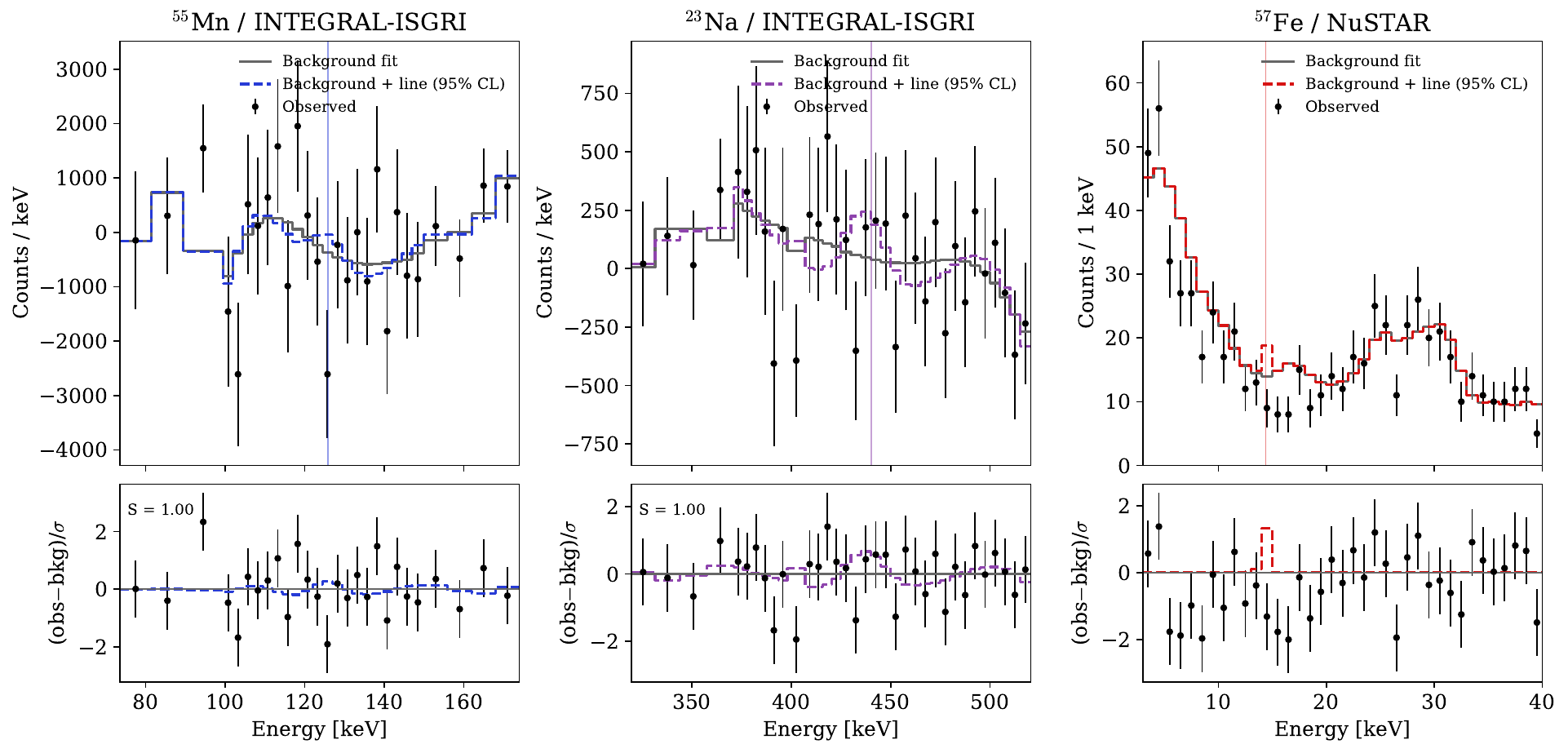}
    \caption{Spectral data (points) around the $^{55}$Mn and $^{23}$Na lines (INTEGRAL-ISGRI) and the $^{57}$Fe line (NuSTAR, both modules, 1~keV bins for display; the fit uses the native 0.04~keV channels), with the
    background-only fit (solid) and the fit with the line at its 95\%~CL limit (dashed). The ISGRI spectra are the mask-deconvolved spectra of Betelgeuse from the 375 pointings with Betelgeuse in the fully coded field of view (68 revolutions), combined with the weights of the fit and shown as source counts per keV for the total exposure of 0.735~Ms, in the OSA energy channels (2.5~keV wide at 100--150~keV and 5~keV wide at 370--520~keV); the background is the residual continuum left after the mask deconvolution. Lower panels: residuals in units of the uncertainty $(\sigma)$ used in the fit (for ISGRI including the error scaling $s_r$ and the scale factor $S$), and the 95\%~CL line as seen by the fit.}
    \label{fig:conversion_spectra}
  \end{figure*}
We use data from the Imager on-Board the INTEGRAL Satellite (IBIS), specifically its low-energy detector layer, the INTEGRAL Soft Gamma-Ray Imager (ISGRI), a coded-mask instrument with CdTe detectors operating in the 15~keV--1~MeV range. This energy range makes INTEGRAL-ISGRI well suited to search for the $^{23}$Na signal, through both magnetic conversion and decay, and the $^{55}$Mn signal, through magnetic conversion. IBIS has a fully coded field of view of $9^\circ\times9^\circ$ and an angular resolution of $12'$ FWHM~\cite{2003A&A...411L...1W}. We use all 377 science windows in which Betelgeuse lies within the fully coded field of view, from 68 INTEGRAL revolutions between 2003 (revolution 0045) and 2024 (revolution 2818); 375 of them could be processed, with a total exposure of 0.735~Ms. For each science window we extract the spectrum of Betelgeuse with OSA, which separates it from the detector background and the other sources in the field by the coded-mask pattern, and take the effective area from the OSA response matrix of that science window. The expected count rates are obtained by forward-folding the ALP-induced photon spectrum through this response matrix, which includes the effective area and the energy resolution of ISGRI.
\subsection{Analysis procedure}
\label{LIKELIHOOD_ANALYSIS}
We perform a likelihood analysis of NuSTAR and INTEGRAL data shown in Fig.~\ref{fig:conversion_spectra}. The two instruments require different treatments of the background: NuSTAR is a focusing telescope, so the background can be measured in a source-free region of the same observation, whereas ISGRI is a coded-mask imager whose whole detector is illuminated by the source, so the background must be modelled from the measured spectrum itself. \paragraph{NuSTAR.}
 For each NuSTAR module $M \in \{A,B\}$ we use the likelihood \cite{Cand_n_2025},
\begin{equation}
          \mathcal{L}
          = \prod_{M\in\{A,B\}}
          \left[
          \mathcal{G}\!\left(\delta_M,\sigma_{\mathrm{bkg}}\right)
          \prod_j
          \mathrm{Poisson}\!\left(
          N^{j}_{{\rm obs},M}
          \middle|
          N^{j}_{{\rm exp},M}
          \right)
          \right],
          \label{eq:combined_like}
      \end{equation}
      with                                                                
      \begin{equation}
          N^{j}_{{\rm exp},M}
          =
          N^{j}_{{\rm ax},M}
          +
          N^{j}_{{\rm bkg},M}
          \left(1+\delta_M\right).
      \end{equation}
    Here, $N^{j}_{{\rm obs},M}$, $N^{j}_{{\rm ax},M}$, and $N^{j}_{{\rm bkg},M}$ denote the observed, expected signal, and nominal background counts in energy channel $j$ of module $M$. We use the native 0.04~keV channels of NuSTAR, which makes Eq.~\eqref{eq:combined_like} equivalent to the unbinned likelihood of Ref.~\cite{Cand_n_2025}. The nominal background is the spectrum of a source-free region, smoothed with a Gaussian of 1~keV standard deviation and rescaled to the area of the source region. The background normalization $\delta_{M}$ of each module is a nuisance parameter,
    constrained by the Gaussian term $\mathcal{G}$ with $\sigma_{\rm bkg}=10\%$ for both FPMA and FPMB~\cite{Xiao_Carenza_Giannotti_Mirizzi_Perez_Straniero_Grefenstette}, and is profiled.
 \paragraph{INTEGRAL-ISGRI.}
INTEGRAL observes in revolutions, \textit{i.e.}\ orbits, which lasted 72~hr until early 2015 and 64~hr since~\cite{Kuulkers:2021svd}. During a revolution the satellite points at a fixed sky direction for
30--60~min, then shifts its pointing by $2.17^\circ$ in a regular dither pattern, and repeats~\cite{Kuulkers:2021svd}. Each fixed pointing is processed as one unit, called a science
window~\cite{Courvoisier:2003vm}. We use all science windows with Betelgeuse in the fully coded field of view, \textit{i.e.}\ less than $4.5^\circ$ off-axis: 375 pointings from 68 revolutions between 2003
and 2024, with a total exposure of $0.735$~Ms. We process them with OSA, which reconstructs the photon energies with the current calibration and, in each energy channel, fits the shadow patterns of all catalogued sources in the field, with Betelgeuse added at its fixed position, together with the detector background. Science windows of the same revolution share the energy calibration and background level, while both change between revolutions. Within each revolution we combine the spectra of its pointings channel by channel, weighted by their inverse variances. Each revolution $r$ therefore has its own background model, while the ALP coupling is common to all revolutions. The mask deconvolution removes the detector background, and what remains is a smooth residual continuum, which we model by a cubic spline in the count rate with knots spaced by $1.5$ times the local FWHM of a line in the OSA response. The spline coefficients are profiled for each tested value of $(m_a,g)$. For magnetic conversion we fit each line in a window of $\pm3$ FWHM around it ($77$--$175$~keV for $^{55}$Mn and $323$--$520$~keV for $^{23}$Na); for decay we fit $30$--$450$~keV, since below about 27~keV most pointings have no data. The extracted rates have Gaussian uncertainties, so the likelihood of each revolution is Gaussian in the rates, and the total
likelihood is the product over revolutions. If the background-only fit of a revolution has $\chi^2/{\rm dof}>1$, we scale its uncertainties by a common factor $s_r$ chosen such that $\chi^2/{\rm dof}=1$;
this gives $s_r\leq1.39$. For each revolution $r$ we fit the spectrum with the background and the amplitude $a$ of the ALP line, $a\propto(g_{a\gamma}g_{aN})^2$, which gives the best-fit amplitude and its uncertainty, $\hat a_r\pm\sigma_r$. Since Betelgeuse is a steady source, $a$ is common to all revolutions, and we combine the revolutions following the unconstrained averaging procedure of the Particle Data Group (Ref.~\cite{ParticleDataGroup:2024cfk}, Introduction, Sec.~5.2.2).
The weighted average and its uncertainty are [Ref.~\cite{ParticleDataGroup:2024cfk}, Eq.~(1)]
\begin{equation}
    \bar a\pm\sigma=\frac{\sum_r w_r\,\hat a_r}{\sum_r w_r}\pm\Bigl(\sum_r w_r\Bigr)^{-1/2},
    \qquad w_r=\sigma_r^{-2},
\end{equation}
which is the joint fit of all revolutions with a single common amplitude. We then compute $\chi^2=\sum_r w_r(\bar a-\hat a_r)^2$. If the uncertainties $\sigma_r$ are correct, its expectation value is the number of degrees of freedom, $N_{\rm rev}-1$. If $\chi^2/(N_{\rm rev}-1)\leq1$, the revolutions are consistent and $\sigma$ is used as it is. If $\chi^2/(N_{\rm rev}-1)>1$, the revolutions disagree by more than their uncertainties allow, most likely because these are underestimated; not knowing which revolution is responsible, we assume that all are underestimated by the same factor and multiply $\sigma$ by [Ref.~\cite{ParticleDataGroup:2024cfk}, Eq.~(2)]
\begin{equation}
    S=\left[\frac{\chi^2}{N_{\rm rev}-1}\right]^{1/2},
  \end{equation}
which is equivalent to dividing $\Delta\ln\mathcal{L}$ by $S^2$. We find $S\leq1.04$ for all considered isotopes, so the revolutions agree within their uncertainties.
\paragraph{Magnetic conversion.}
The converted photon carries the energy of the ALP, so the signal is the stellar line folded
through the detector response~\cite{Cand_n_2025},
\begin{equation}
  \begin{aligned}
  N^{j}_{{\rm ax},M} ={}& \frac{\Delta t_M}{4\pi D^2}
  \int {\rm d}E_a\;
  R_M(E_j,E_a)\,A_{{\rm eff},M}(E_a)
  \\
  &\times
  \frac{{\rm d}\dot N_a}{{\rm d}E_a}(g_{aN}^{\rm eff})\,
  P_{a\to\gamma}(E_a,g_{a\gamma}) ,
  \end{aligned}
  \label{eq:nsig-primakoff}
  \end{equation}
where $\Delta t_M$ and $A_{{\rm eff},M}(E_a)$ are the exposure time and effective area of NuSTAR
module $M$ (for ISGRI, $M$ is replaced by the revolution $r$), and $R_M(E_j,E_a)$ is the energy
redistribution function, i.e.\ the probability that a photon of energy $E_a$ is recorded in channel
$j$. For NuSTAR, $A_{\rm eff}$ and $R$ are taken from the ancillary response and response matrix
files generated by \texttt{nuproducts} for the source region of each module. For ISGRI, $A_{{\rm eff}}$ and $R$ are given by the OSA response matrices of the science windows in revolution $r$, which include both the effective area and the energy resolution of ISGRI.
${\rm d}\dot N_a/{\rm d}E_a$ is the total ALP production rate, Eq.~\eqref{eq:production_rate}, $D$ is
the distance to Betelgeuse, and $P_{a\to\gamma}$ is the ALP-to-photon conversion probability.
\paragraph{ALP decay.}
For decay, the expected counts in channel $j$ of module $M$ (revolution $r$ for ISGRI)
are~\cite{Candon:2024eah},
\begin{equation}
    \begin{aligned}
    N^{j}_{{\rm ax},M} ={}& \frac{\Delta t_M\,\dot N_a(g^{\rm eff}_{aN})}{4\pi D^2}
    \left(\frac{p_*}{E_*}\right)^3
    \Biggl\langle \int_{(E_a-p_a)/2}^{(E_a+p_a)/2} {\rm d}E_\gamma\,\frac{2}{p_a}
    \\
    &\times P_{a\to\gamma\gamma}(g_{a\gamma})\,R_M(E_j,E_\gamma)\,A_{{\rm eff},M}(E_\gamma)
    \Biggr\rangle_{v_\parallel},
    \end{aligned}
    \label{eq:nsig_decay}
  \end{equation}
where $(p_*/E_*)^3$, with $p_*=\sqrt{E_*^2-m_a^2}$, is the phase-space suppression of the
emission of a massive ALP in the rest frame of the nucleus, and $R_M$ and $A_{{\rm eff},M}$ are the energy redistribution function and effective area defined below Eq.~\eqref{eq:nsig-primakoff}. For a nucleus velocity component $v_\parallel$ along the ALP direction,
\begin{equation}
  p_a^2=\bigl(p_*+E_*v_\parallel\bigr)^2+2\sigma_{\rm eff}^2,
  \qquad
  E_a^2=m_a^2+p_a^2,
  \label{eq:thermal_boost}
  \end{equation}
where $v_\parallel$ is Gaussian with width $\sigma_{\rm eff}/E_*$, $\sigma_{\rm eff}$ is the width of the line profile, and the transverse velocity enters only through its mean square. The average over $v_\parallel$ thus replaces ${\rm d}\dot N_a/{\rm d}E_a$ of Eq.~\eqref{eq:nsig-primakoff}, to which it reduces for $m_a\ll E_*$.
\paragraph{Statistical treatment.}
For a given ALP mass we compare each signal hypothesis with the best fit to the data, refitting the background parameters $\theta$ ($\delta_M$ for NuSTAR, the spline coefficients for ISGRI) for every value of the coupling $g$,
\begin{equation}
  \mathcal{L}_{\rm p}(g)=\max_{\theta}\,\mathcal{L}(g,\theta),
  \qquad
  q(g)=-2\ln\frac{\mathcal{L}_{\rm p}(g)}{\mathcal{L}_{\rm p}(\hat g)}\quad (g>\hat g),
\end{equation}
where $\hat g$ is the best fit, whose signal is allowed to be negative. This is the one-sided test
statistic $q_\mu$ of Ref.~\cite{Cowan:2010js}, which asymptotically follows a half-$\chi^2$ distribution with one degree of freedom, so that couplings with $q>2.71$ are excluded at 95\% CL. For ISGRI, $q$ is divided by $S^2$. Where the data fluctuate more than $1\sigma$ below the background, we replace the limit by the $-1\sigma$ expected limit (power constraint~\cite{Cowan:2011an}). In the decay channel we evaluate $q$ at the lower edge of the band, where the signal is proportional to $g_{a\gamma}^2$; at the upper edge the signal falls exponentially with $g_{a\gamma}$, so the edge is insensitive to the choice of reference.

\section{RESULTS}
\label{Results}
 We find no significant excess in either the INTEGRAL-ISGRI or the NuSTAR data. We therefore derive 95\% CL upper limits on the ALP coupling for the two detection channels: magnetic conversion ($a\to\gamma$) and decay ($a\to\gamma\gamma$). Each transition constrains a different combination of the nucleon couplings, $g_{aN}^{\rm eff}=c_p\,g_{ap}+c_n\,g_{an}$, set by its nuclear matrix element; consequently, limits on $|g_{a\gamma}g_{aN}^{\rm eff}|$ for different isotopes cannot be shown on the same axes. We therefore show them separately for $^{23}$Na and $^{55}$Mn (Figs.~\ref{fig:primakoff_final} and~\ref{fig:decay_ganeff}), and compare all three isotopes along with existing constraints in the limiting cases $g_{an}=0$ and $g_{ap}=0$ (Figs.~\ref{fig:primakoff_proton} and~\ref{fig:decay_final}). 
 
 The main results are: (i) $^{23}$Na gives the strongest limits in both channels. Although its transition energy is higher, it is comparatively more abundant at the end of core carbon burning; (ii) for the decay channel, $^{23}$Na improves on the leading astrophysical constraint (combined M82 and neutron-star cooling bound) on $g_{ap}$ for $m_a\simeq155$--$440$~keV, whereas for $g_{an}$, the leading astrophysical constraint remains dominant;  (iii) for the magnetic conversion channel, the $^{23}$Na constraint on $g_{ap}$ surpasses the previously published $^{57}$Fe constraint by two orders of magnitude~\cite{Cand_n_2025}, whereas for $g_{an}$ they are comparable ($^{23}$Na yields better limits only for $m_a\gtrsim 3 \times 10^{-11}$ eV); (iv) for $g_{ap}$, the $^{55}$Mn constraint is comparable to the previously published $^{57}$Fe bound at lower masses~\cite{Cand_n_2025}, yet surpasses it at higher masses for both the magnetic conversion and decay channels, owing to its higher transition energy. All limits shown here assume the fiducial model at the end of core carbon burning, and their dependence on the evolutionary stage is discussed in App.~\ref{uncertainty}. In the following, we qualitatively discuss the notable features of the exclusion plots.
\subsection{Constraints from ALP magnetic conversion}
All magnetic conversion limits on the ALP coupling reach a mass-independent plateau for ultralight ALP masses, where the coherence factor of the conversion probability saturates, \textit{i.e.}, $\sin^2(\Delta_{\rm osc}D/2)/(\Delta_{\rm osc}D/2)^2\to1$. In this regime, the ALP-photon oscillation length greatly exceeds the distance to Earth, and the conversion probability depends only on the product $g_{a\gamma}B_T D$, rendering the limit independent of $m_a$. For $m_a\gg\omega_{\rm pl}$, the oscillation wavenumber grows as $\Delta_{\rm osc}\propto m_a^2/E_a$, so the coherence is progressively lost as the mass increases. The limits begin to oscillate once $\Delta_{\rm osc}D/2\sim1$, marking the transition from the coherent to the incoherent regime. Since $\Delta_{\rm osc}\propto 1/E_a$, isotopes with higher transition energy have a smaller oscillation wavenumber at a given $m_a$, and therefore maintain coherence up to higher masses. This transition occurs at $m_a\simeq4.3\times10^{-11}$~eV for $^{57}$Fe ($E_*=14.4$~keV), $1.3\times10^{-10}$~eV for $^{55}$Mn ($E_*=125.9$~keV), and $2.4\times10^{-10}$~eV for $^{23}$Na ($E_*=440.2$~keV).

The limits presented in this analysis are also compared with two leading astrophysical bounds (light grey shaded): (a) the recent SN~1987A bound derived from the magnetic conversion of axions, produced in the stellar core, in the Galactic magnetic field~\cite{Fiorillo:2025gnd,Fiorillo:2026rgo, Ning:2025kyu}, and (b) the combined bound from the NuSTAR X-ray searches in M82 on $g_{a\gamma}$~\cite{Ning:2024eky} and the neutron-star cooling bound on $g_{ap}$ or $g_{an}$~\cite{Buschmann:2021juv}.

\subsubsection{ALP--effective nucleon coupling}
\begin{figure*}[t]
    \centering

    \begin{minipage}[t]{0.49\textwidth}
        \centering
        \includegraphics[width=\linewidth]{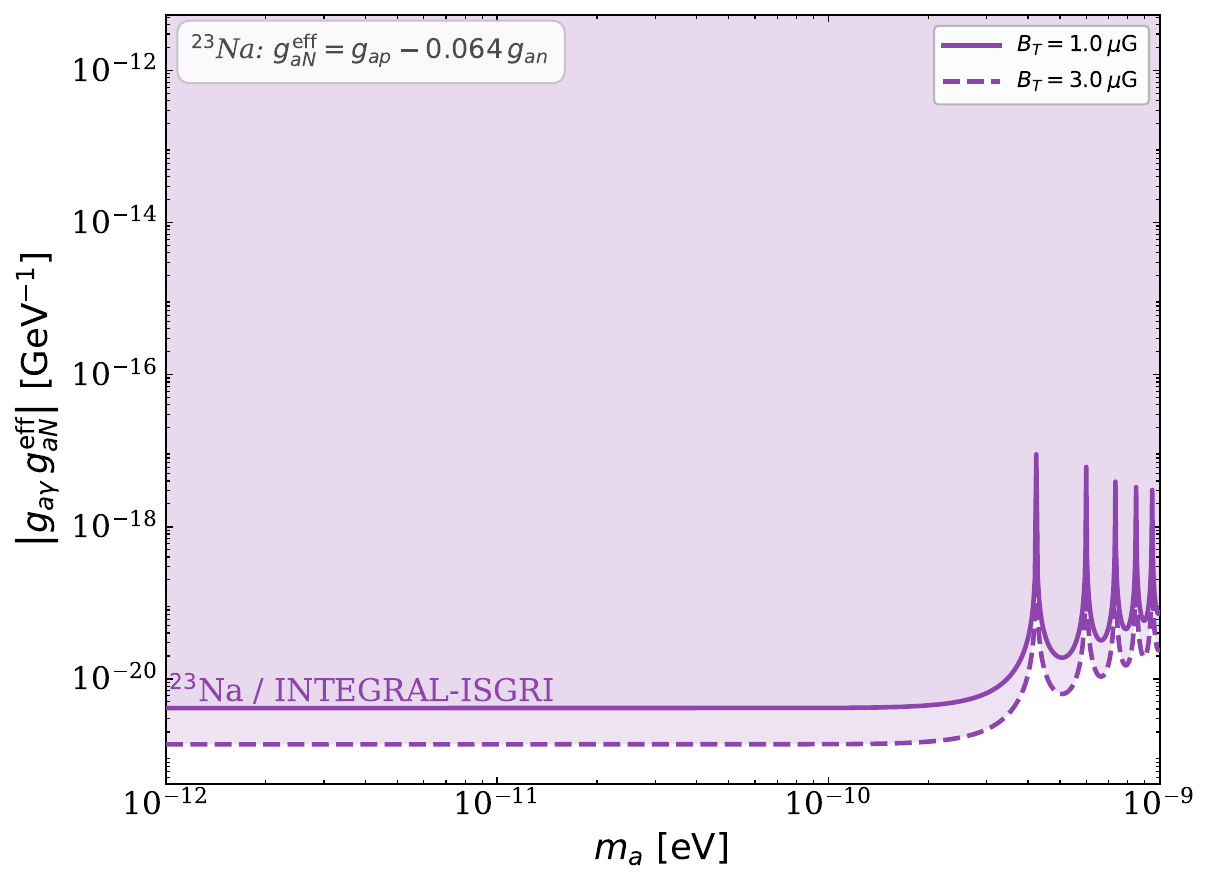}
    \end{minipage}
    \hfill
    \begin{minipage}[t]{0.49\textwidth}
        \centering
        \includegraphics[width=\linewidth]{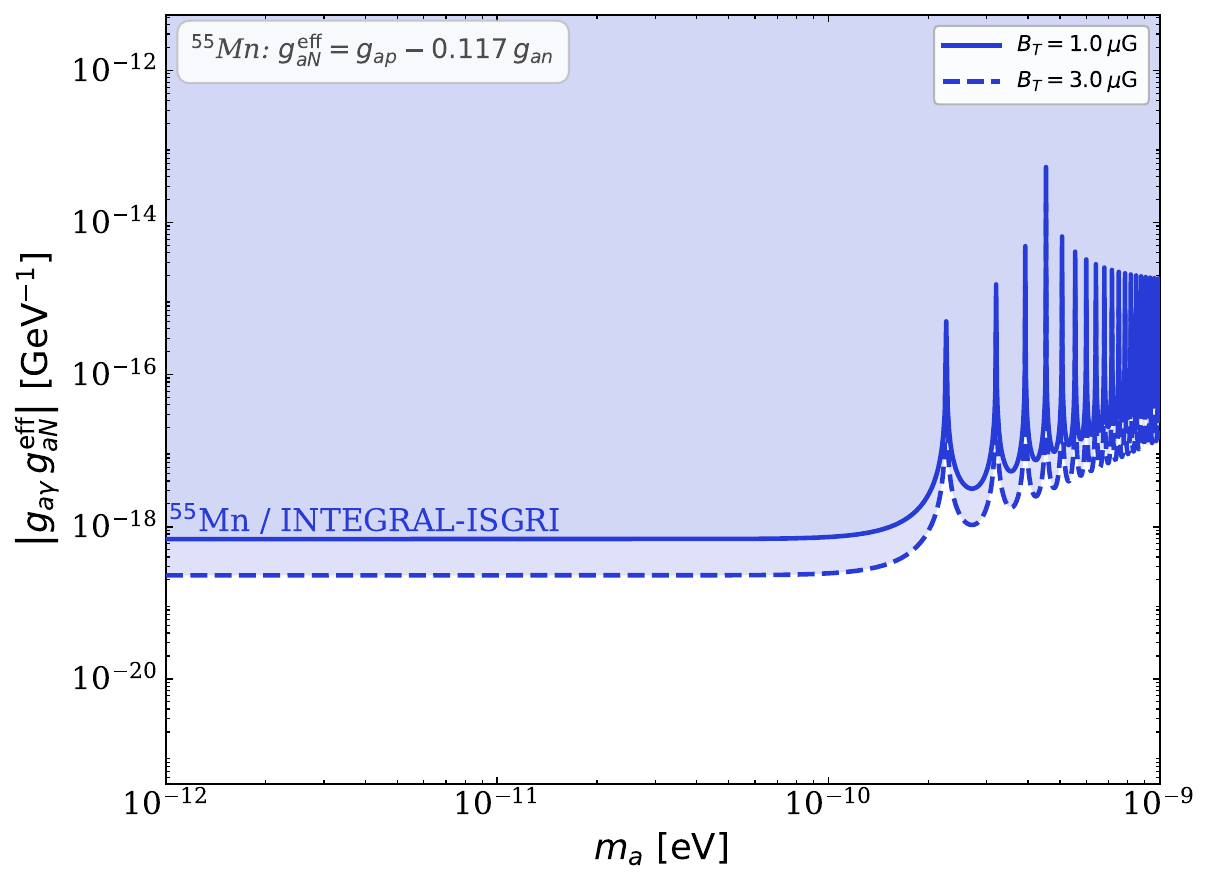}
    \end{minipage}

    \caption{95\% CL exclusion limits from ALP magnetic conversion on $|g_{a\gamma}g_{aN}^{\rm eff}|$ for $^{23}$Na/INTEGRAL-ISGRI (purple) and $^{55}$Mn/INTEGRAL-ISGRI (blue), for the fiducial Betelgeuse model at the end of core carbon burning ($T_c=95$~keV). Solid lines correspond to a Galactic magnetic field strength $B_T=1\,\mu$G and dashed lines to $B_T=3\,\mu$G.}

    \label{fig:primakoff_final}
\end{figure*}
Fig.~\ref{fig:primakoff_final} shows the exclusion limits on $|g_{a\gamma}g_{aN}^{\rm eff}|$ obtained from this analysis. For $^{23}$Na, the effective nucleon coupling is $(g_{ap} - 0.064 g_{an})$ whereas, for $^{55}$Mn, it corresponds to $(g_{ap} - 0.117 g_{an})$. In the ultralight mass regime, the limits approach a mass-independent plateau, with values of $4.1\times10^{-21}\,{\rm GeV^{-1}}$ and $6.9\times10^{-19}\,{\rm GeV^{-1}}$ for $^{23}$Na and $^{55}$Mn, respectively. At higher masses, the limits develop an oscillatory feature, arising from the interference between the ALP and photon propagation phases accumulated over the coherence length of the Galactic magnetic field. This oscillation sets in at $m_a\simeq2.4\times10^{-10}$~eV for $^{23}$Na and at $m_a\simeq1.3\times10^{-10}$~eV for $^{55}$Mn, with the later onset for $^{23}$Na reflecting its higher transition energy and correspondingly more energetic ALPs.

\subsubsection{ALP--proton coupling}
\begin{figure*}[t]
      \centering
      \begin{minipage}[t]{0.49\textwidth}
          \centering
          \includegraphics[width=\linewidth]{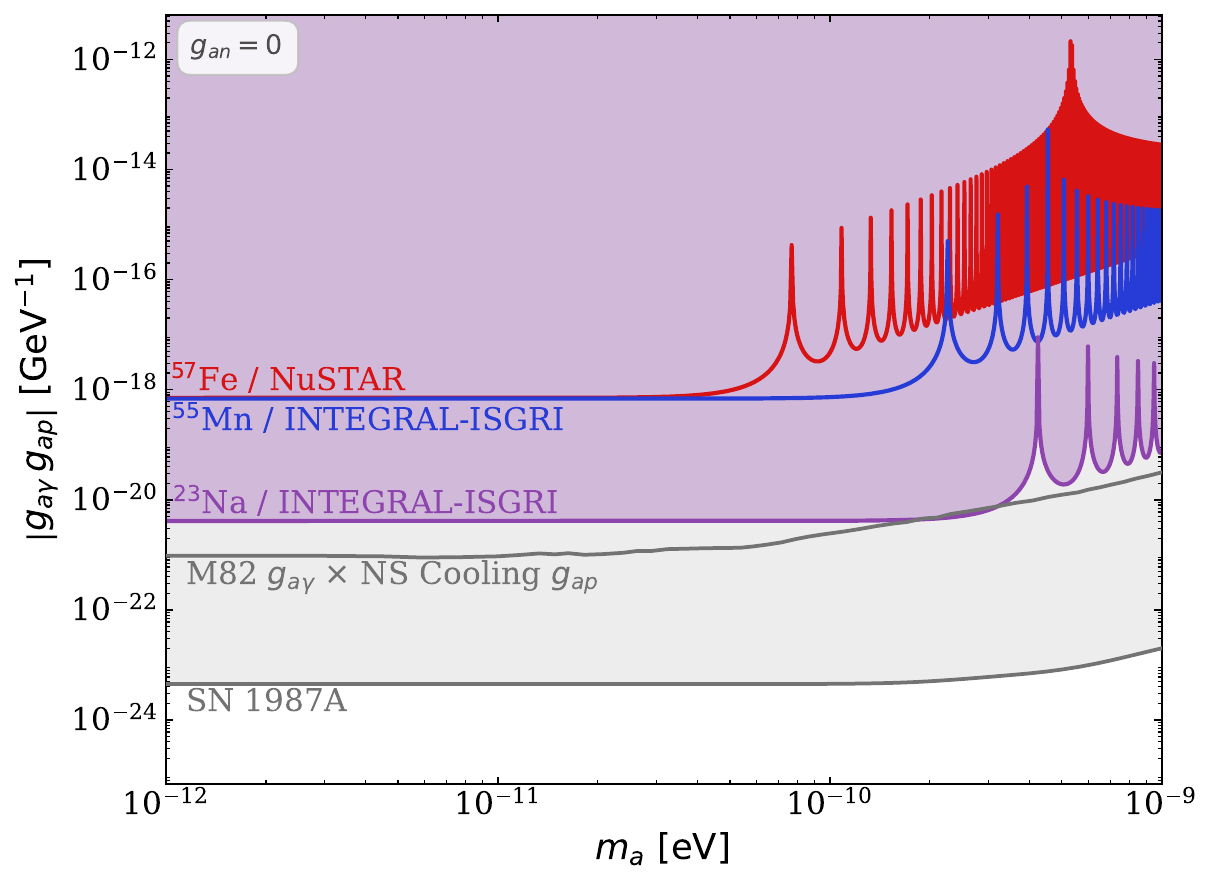}
      \end{minipage}
      \hfill
      \begin{minipage}[t]{0.49\textwidth}
          \centering
          \includegraphics[width=\linewidth]{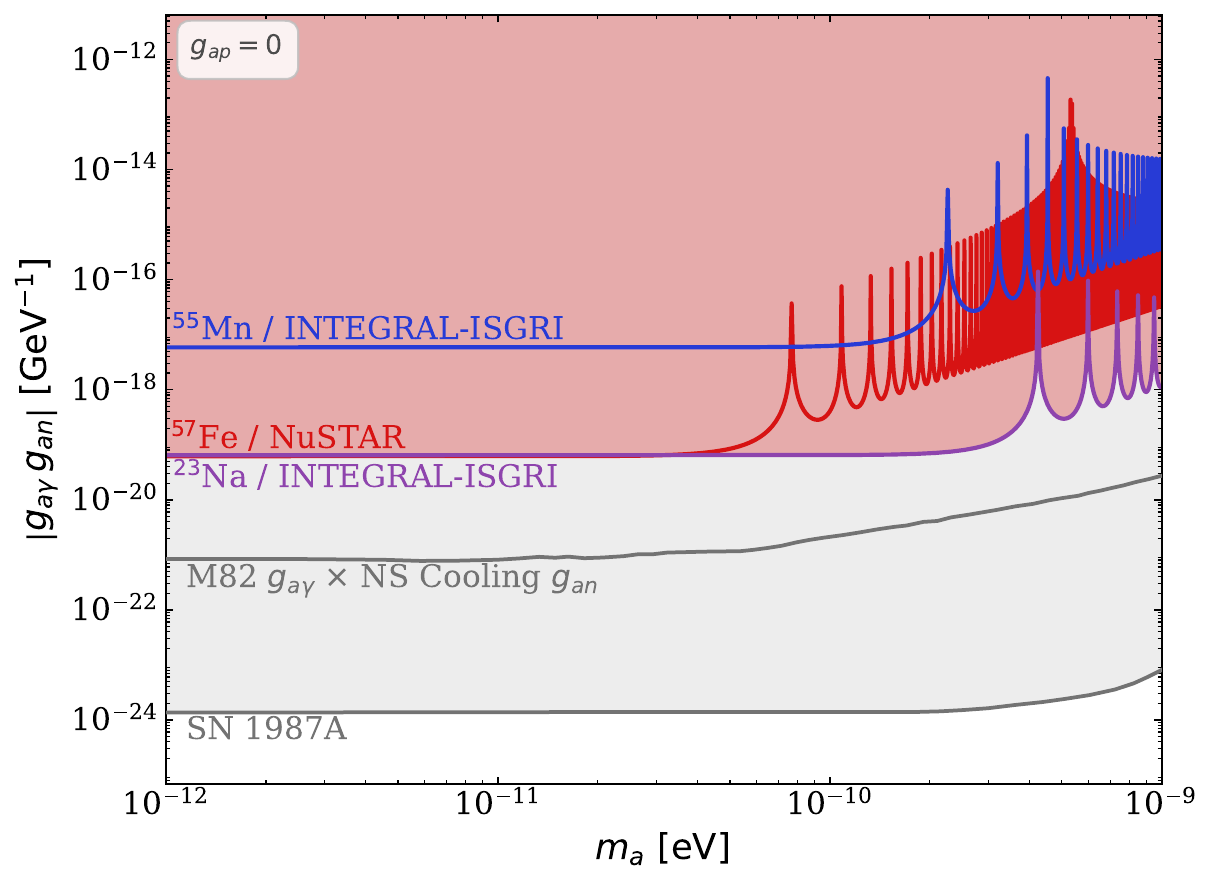}
      \end{minipage}
\caption{95\% CL exclusion limits from ALP magnetic conversion on $|g_{a\gamma}g_{ap}|$ (left, $g_{an}=0$) and $|g_{a\gamma}g_{an}|$ (right, $g_{ap}=0$) for $^{23}$Na/INTEGRAL-ISGRI (purple),
$^{55}$Mn/INTEGRAL-ISGRI (blue) and $^{57}$Fe/NuSTAR (red), for the fiducial Betelgeuse model at the end of core carbon burning ($T_c=95$~keV) and a Galactic magnetic field $B_T=1\,\mu$G. The coloured areas are excluded by our analysis. The grey curves are existing bounds on the same coupling products: (a) the product of the M82 bound on $g_{a\gamma}$~\cite{Ning:2024eky} and the neutron-star cooling bound on $g_{ap}$ or $g_{an}$~\cite{Buschmann:2021juv}; and (b) the SN~1987A bound, derived from the conversion of supernova axions in the Galactic magnetic field for the $g_{ap}$ case (Ref.~\cite{Fiorillo:2025gnd}) and from Ref.~\cite{Ning:2025kyu} for the $g_{an}$ case.}
    \label{fig:primakoff_proton}
  \end{figure*}
For the ALP-proton coupling ($g_{an}=0$, Fig.~\ref{fig:primakoff_proton}, left), $^{23}$Na yields the strongest limit among all three isotopes considered here. Quantitatively, in the ultralight mass regime, where the constraints become mass-independent, we obtain $|g_{a\gamma}g_{ap}|<4.1\times10^{-21}\,{\rm GeV^{-1}}$ for $^{23}$Na. This is two orders of magnitude stronger than the $^{55}$Mn and previously published $^{57}$Fe-based limits, which are $|g_{a\gamma}g_{ap}|<6.9\times10^{-19}$ and $7.1\times10^{-19}\,{\rm GeV^{-1}}$, respectively. The oscillation feature emerges at relatively higher ALP masses for $^{23}$Na, owing to its higher transition energy of 440.2~keV compared to $^{55}$Mn (125.9~keV) and $^{57}$Fe (14.4~keV). Note that all these regions are already excluded by the recent SN~1987A analysis~\cite{Fiorillo:2025gnd,Fiorillo:2026rgo}, and our constraints therefore do not yield new exclusions in this regime. Indeed, the SN~1987A bound, $|g_{a\gamma}g_{ap}|<4.5\times10^{-24}\,{\rm GeV^{-1}}$ for $m_a\lesssim10^{-10}$~eV, is at least a factor of $\simeq900$ stronger than our $^{23}$Na-based limit across the entire mass range shown. The combined M82 and neutron-star cooling bound, $|g_{a\gamma}g_{ap}|<9.6\times10^{-22}\,{\rm GeV^{-1}}$ in the massless limit, is comparatively closer, lying only a factor of $\simeq4$ below our $^{23}$Na-based limit and nearly overlapping with it.
\subsubsection{ALP--neutron coupling}
For the ALP-neutron coupling ($g_{ap}=0$, Fig.~\ref{fig:primakoff_proton}, right), the $^{57}$Fe and $^{23}$Na-based limits yield comparable plateaus of $|g_{a\gamma}g_{an}|<6.2\times10^{-20}$ and $6.4\times10^{-20}\,{\rm GeV^{-1}}$, respectively. This near-equivalence arises from a compensation between two competing factors: $^{57}$Fe benefits from its neutron-dominated nuclear matrix element, whereas $^{23}$Na compensates through its higher abundance at the end of core carbon burning. $^{55}$Mn-based limits for the ALP-neutron coupling are significantly weaker than $^{57}$Fe and $^{23}$Na-based limits. Similar to the $g_{ap}$ case, the oscillation feature emerges at relatively higher ALP masses for $^{23}$Na, owing to its higher transition energy, while $^{57}$Fe exhibits the oscillation feature at the lowest ALP masses among the three isotopes. The leading astrophysical bounds on $g_{an}$ are considerably stronger than the results obtained in this analysis, and our limits therefore do not yield new exclusions. Quantitatively, the combined M82 and neutron-star cooling bound, $|g_{a\gamma}g_{an}|<8.3\times10^{-22}\,{\rm GeV^{-1}}$ in the massless limit, lies a factor of $\simeq75$ below the $^{57}$Fe and $^{23}$Na plateaus, while the SN~1987A bound, $|g_{a\gamma}g_{an}|\lesssim1.4\times10^{-24}\,{\rm GeV^{-1}}$, is more than four orders of magnitude below our $^{57}$Fe and $^{23}$Na plateaus~\cite{Ning:2025kyu}.

\subsection{Constraints from ALP decay}
For the ALP decay channel, our results exclude a band in the coupling space for each isotope. The upper edge of the band is determined by the condition that ALPs decay inside Betelgeuse and the resulting photons are absorbed before escaping. The lower edge is set by the condition that the decay length $L_a$ greatly exceeds the distance to Earth, so that only a small fraction $\propto D/L_a$ of the ALPs decay along the line of sight. Both edges follow from comparing $L_a\propto1/(g_{a\gamma}^2m_a^4)$ with a fixed length scale — the distance to Earth for the lower edge and the stellar radius for the upper — so both scale as $g_{a\gamma}\propto m_a^{-2}$ and run roughly parallel. For NuSTAR, and for ISGRI above $m_a\sim1$~keV, the lower edge falls more slowly as $\propto m_a^{-3/2}$: both instruments collect only those decay photons that arrive within a small angular aperture of Betelgeuse ($\beta_{\rm ap}=60''$ for NuSTAR and $6'$ for the ISGRI point-source spectra), and this fraction decreases as heavier ALPs are less boosted. As $m_a\to E_*$, the phase-space factor $(p_*/E_*)^3$ suppresses the emission and the band closes just below $E_*$. Near the high-mass end of each band, the lower edge exhibits small oscillations. These are statistical fluctuations of the data. In this regime, the expected signal is concentrated in a few bins at the low-energy edge of the decay box, $E_\gamma=(E_a-p_a)/2$, so the limit closely follows the counts in those bins as the edge shifts with $m_a$. The largest downward fluctuations are removed by the power constraint~\cite{Cowan:2011an}.

For comparison, we also show the leading astrophysical constraint obtained by combining the M82 decay bound on $g_{a\gamma}$~\cite{Candon:2024eah} with the neutron-star cooling bound on $g_{ap}$ or $g_{an}$~\cite{Buschmann:2021juv}.

\subsubsection{ALP--effective nucleon coupling}
  \begin{figure*}[t]
      \centering
      \begin{minipage}[t]{0.49\textwidth}
          \centering
          \includegraphics[width=\linewidth]{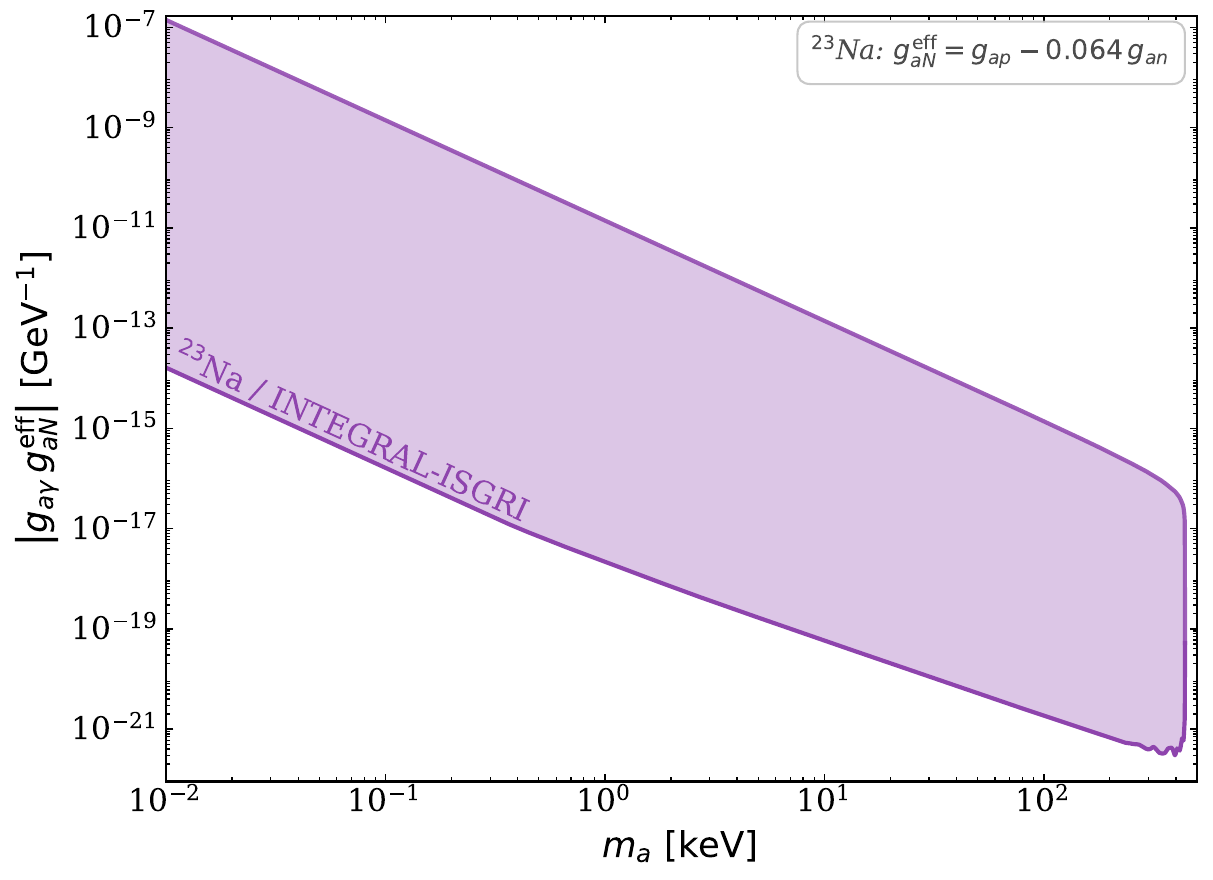}
      \end{minipage}
      \hfill
      \begin{minipage}[t]{0.49\textwidth}
          \centering
          \includegraphics[width=\linewidth]{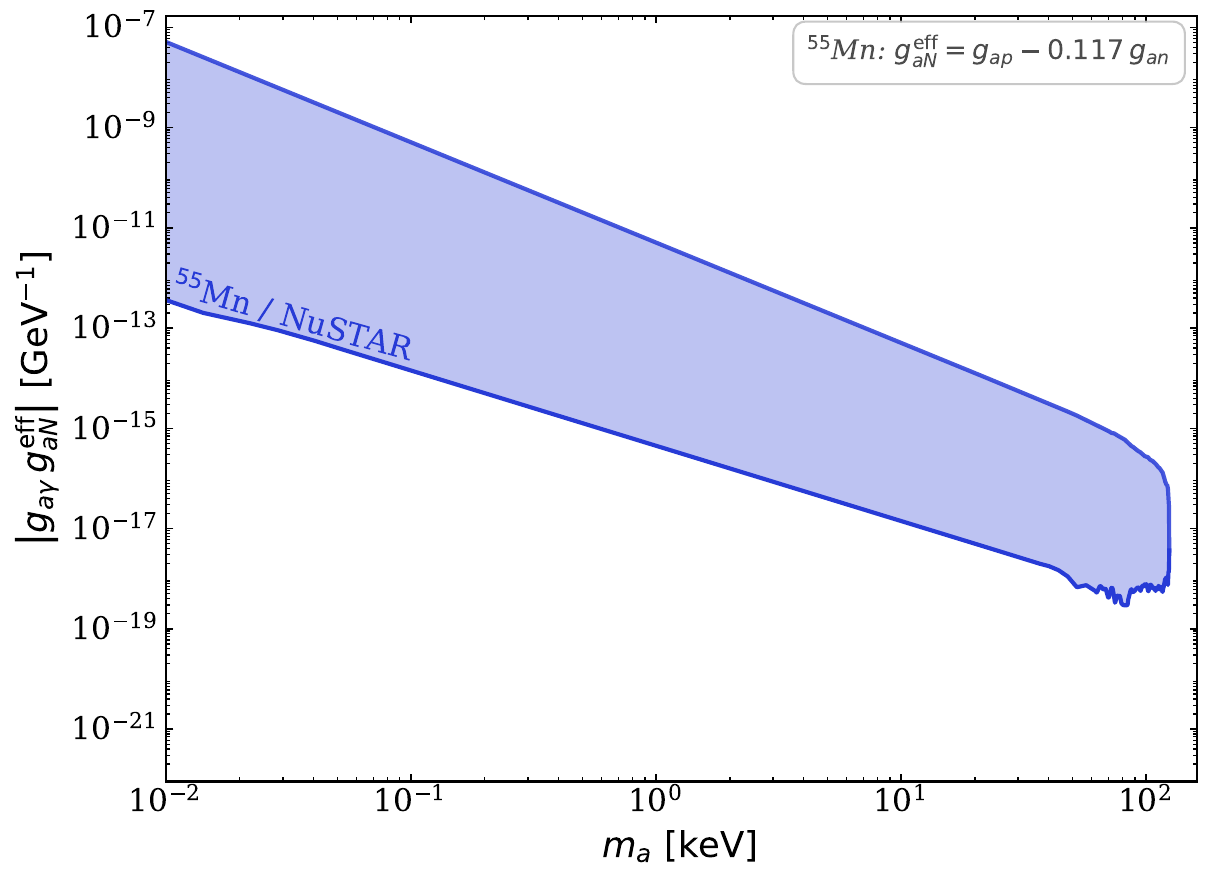}
      \end{minipage}
      \caption{95\% CL exclusion limits from ALP decay ($a\to\gamma\gamma$) on
      $|g_{a\gamma}g_{aN}^{\rm eff}|$ for $^{23}$Na/INTEGRAL-ISGRI (left, purple) and
      $^{55}$Mn/NuSTAR (right, blue), for the fiducial Betelgeuse model at the end of core carbon
      burning ($T_c=95$~keV). The shaded areas are excluded by our analysis.}
      \label{fig:decay_ganeff}
  \end{figure*}
  \begin{figure*}
    \centering
    \begin{minipage}[t]{0.49\textwidth}
        \centering
        \includegraphics[width=\linewidth]{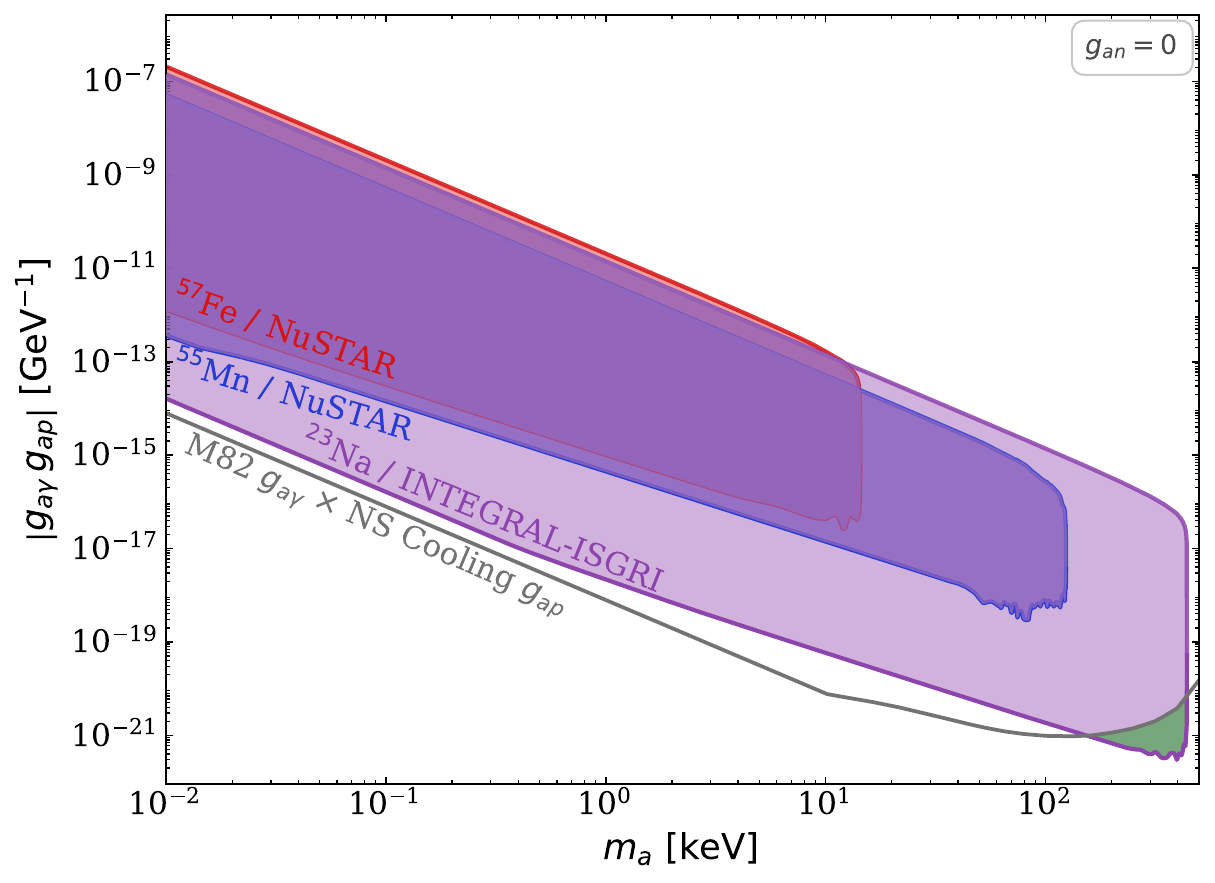}
        \label{fig:decay_final_2}
    \end{minipage}
    \hfill
    \begin{minipage}[t]{0.49\textwidth}
        \centering
        \includegraphics[width=\linewidth]{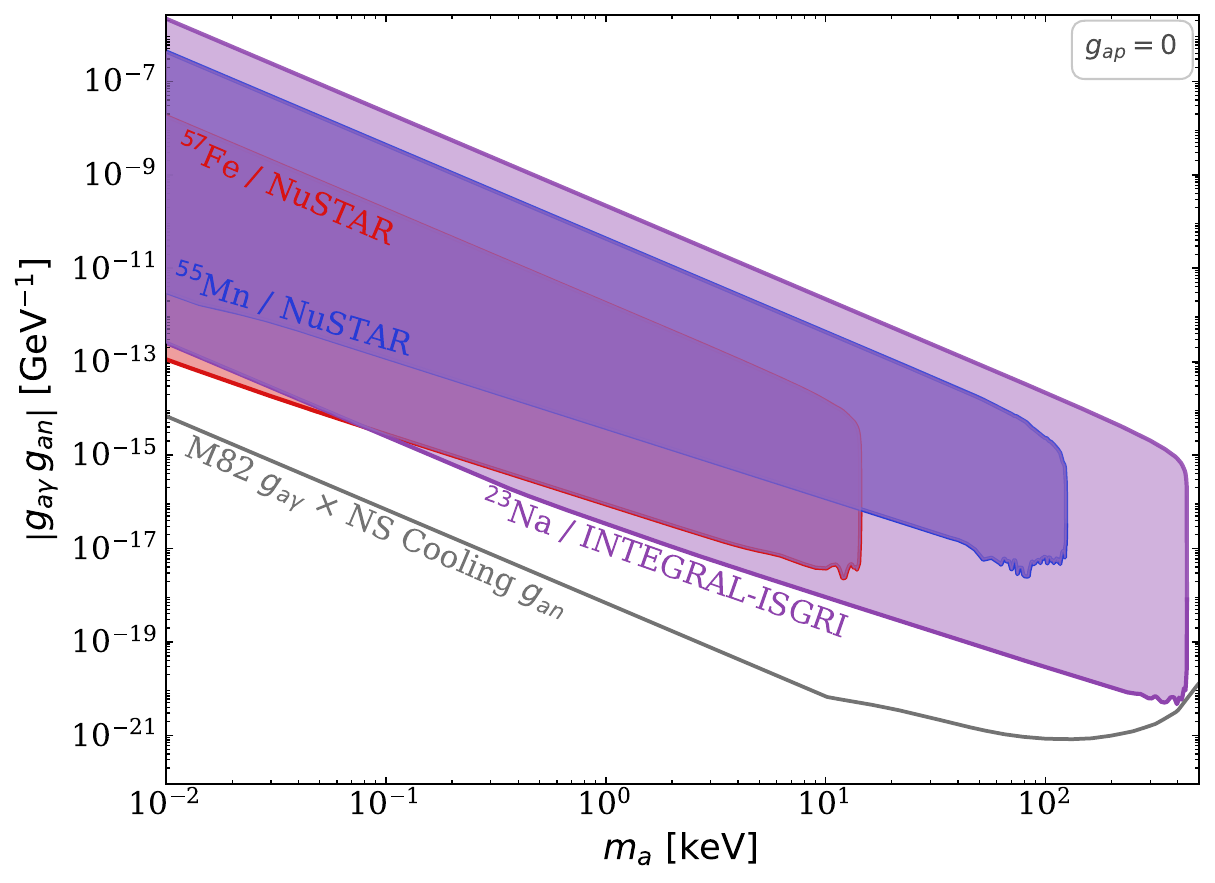}
        \label{fig:decay_final_3}
    \end{minipage}

    \caption{95\% CL exclusion limits from ALP decay ($a\to\gamma\gamma$) on $|g_{a\gamma}g_{ap}|$ (left, $g_{an}=0$) and $|g_{a\gamma}g_{an}|$ (right, $g_{ap}=0$) for $^{23}$Na/INTEGRAL-ISGRI (purple), $^{55}$Mn/NuSTAR (blue) and $^{57}$Fe/NuSTAR (red), for the fiducial Betelgeuse model at the end of core carbon burning ($T_c=95$~keV). The coloured areas are excluded by our analysis. The grey curve is the product of the strongest known bounds on the photon coupling $g_{a\gamma}$ (M82~\cite{Candon:2024eah}) and on the proton or neutron coupling $g_{ap}$, $g_{an}$ (neutron-star cooling~\cite{Buschmann:2021juv}). The green area marks where $^{23}$Na improves on this bound, for $m_a\simeq155$--$440$~keV (left).}

    \label{fig:decay_final}
\end{figure*}
Fig.~\ref{fig:decay_ganeff} shows the decay limits on $|g_{a\gamma}g_{aN}^{\rm eff}|$.  For $^{23}$Na, the effective nucleon coupling is $(g_{ap} - 0.064 g_{an})$ whereas, for $^{55}$Mn, it corresponds to $(g_{ap} - 0.117 g_{an})$. The $^{23}$Na band extends from $1.6\times10^{-14}$ to $1.4\times10^{-7}\,{\rm GeV^{-1}}$ at $m_a=10^{-2}$~keV; its lower edge falls to $1.0\times10^{-21}\,{\rm GeV^{-1}}$ at 150~keV and reaches its minimum of $3.0\times10^{-22}\,{\rm GeV^{-1}}$ near 400~keV, before the band closes at 440.2~keV. The $^{55}$Mn band follows the same shape but is systematically weaker: its lower edge lies a factor of $\sim20$ above that of $^{23}$Na at $10^{-2}$~keV and a factor of $\sim240$ above at 10~keV; it reaches its minimum of $2.9\times10^{-19}\,{\rm GeV^{-1}}$ near 83~keV, and the band closes at 124~keV.

\subsubsection{ALP--proton coupling}
For the ALP-proton coupling ($g_{an} =0$, Fig.~\ref{fig:decay_final}, left), the $^{57}$Fe band reaches $|g_{a\gamma}g_{ap}|\simeq2.6\times10^{-17}\,{\rm GeV^{-1}}$ near 12~keV and closes at its 14.4~keV endpoint. The
$^{55}$Mn band starts a factor of $\sim3.5$ lower at $10^{-2}$~keV, reaches $2.9\times10^{-19}\,{\rm GeV^{-1}}$ near 83~keV, and closes at around 124~keV. $^{23}$Na gives the strongest and widest band, reaching
$3.0\times10^{-22}\,{\rm GeV^{-1}}$ near 400~keV, and more importantly, lies below the leading astrophysical bound for $m_a\simeq155$--$440$~keV. This $^{23}$Na-based decay limit is therefore one of the central results of this work.
\subsubsection{ALP--neutron coupling}
For the ALP-neutron coupling ($g_{ap} = 0$, Fig.~\ref{fig:decay_final}, right), $^{57}$Fe benefits from its neutron-dominated matrix element: its band reaches $|g_{a\gamma}g_{an}|\simeq2.3\times10^{-18}\,{\rm GeV^{-1}}$
near 12~keV and gives the strongest limit below $m_a\simeq0.1$~keV. The $^{55}$Mn band reaches $2.5\times10^{-18}\,{\rm GeV^{-1}}$ near 83~keV and closes at around 124~keV. Above 0.1~keV, $^{23}$Na gives the
strongest limit, reaching $4.7\times10^{-21}\,{\rm GeV^{-1}}$ near 400~keV, and its band closes at 440.2~keV.
\subsection{Summary}
In both channels, \textit{i.e.}, magnetic conversion of ALPs as well as ALP decay, $^{23}$Na/INTEGRAL-ISGRI-based results provide the strongest constraints among the isotopes considered here. For the magnetic conversion channel, it reaches $|g_{a\gamma}g_{ap}|<4.1\times10^{-21}\,{\rm GeV^{-1}}$; however, the recent constraint derived from SN~1987A observations remains more stringent across the entire mass range considered. For the decay channel, $^{23}$Na/INTEGRAL-ISGRI-based results surpass the leading astrophysical constraint on $g_{ap}$ in the ALP mass window of $m_a\simeq155$--$440$~keV.

In the magnetic conversion channel, the $^{23}$Na-based result on $g_{ap}$ improves upon the previously published $^{57}$Fe limit~\cite{Cand_n_2025} by two orders of magnitude and extends the coverage to higher ALP masses, while the $^{55}$Mn-based result is comparable to it in the massless limit and stronger at higher masses. This is attributed to the fact that $^{23}$Na and $^{55}$Mn are odd-proton nuclei, rendering them particularly sensitive to the ALP-proton coupling $g_{ap}$, in contrast to $^{57}$Fe, which is an odd-neutron nucleus. It is worth noting that these conclusions hold under the assumption that Betelgeuse is at the end of core carbon burning; for earlier stages, the $^{23}$Na limits weaken by several orders of magnitude (see Fig.~\ref{fig:magnetic_stellar_uncertainty}).
\section{Conclusions}
\label{conclusion}
We have extended the Betelgeuse ALP search of Ref.~\cite{Cand_n_2025}, which exploited the 14.4~keV nuclear transition of $^{57}$Fe, to two new isotopes: the 125.9~keV transition of $^{55}$Mn and the 440.2~keV transition of $^{23}$Na, using archival INTEGRAL-ISGRI and NuSTAR data. These two isotopes are odd-proton nuclei and therefore provide better sensitivity to the ALP-proton coupling $g_{ap}$, complementing the odd-neutron $^{57}$Fe nucleus, which is primarily sensitive to $g_{an}$. We considered two detection channels for these stellar ALPs: magnetic conversion of ALPs into photons in the Galactic magnetic field ($a\rightarrow\gamma$) and in-flight decay into two photons ($a\rightarrow\gamma\gamma$), the latter being a new channel not considered in Ref.~\cite{Cand_n_2025}.

\paragraph{Magnetic conversion of ALPs.} $^{23}$Na gives the tightest constraint among the three isotopes considered here. For the $g_{ap}$ coupling, the $^{23}$Na-based result yields $|g_{a\gamma}g_{ap}|<4.1\times10^{-21}\,{\rm GeV^{-1}}$, two orders of magnitude stronger than the $^{57}$Fe limit in this plane. Owing to its higher transition energy, the $^{23}$Na constraint also extends to higher ALP masses compared to $^{57}$Fe. For the $g_{an}$ coupling, $^{57}$Fe and $^{23}$Na yield similar limits, as the neutron-dominated nuclear matrix element of $^{57}$Fe is compensated by the higher abundance of $^{23}$Na at the end of core carbon burning. We note, however, that all these regions are already excluded by the recent SN~1987A bound~\cite{Fiorillo:2025gnd,Fiorillo:2026rgo}, which remains more stringent across the entire mass range considered. Furthermore, these conclusions are contingent on the evolutionary stage of Betelgeuse; for earlier evolutionary stages, the $^{23}$Na-based results can weaken substantially.

\paragraph{ALP decay.} Since an isotope can only probe ALPs lighter than its transition energy, the 125.9~keV line of $^{55}$Mn and the 440.2~keV line of $^{23}$Na significantly extend the decay channel sensitivity from $m_a\lesssim14$~keV for $^{57}$Fe to $m_a\lesssim124$~keV and $m_a\lesssim440$~keV, respectively. For the ALP-proton coupling, the $^{55}$Mn and $^{57}$Fe-based results are qualitatively similar, whereas the $^{23}$Na-based results are substantially stronger and surpass the leading astrophysical constraint~\cite{Candon:2024eah, Buschmann:2021juv} in the mass window $m_a\simeq155$--$440$~keV. For the ALP-neutron coupling, the leading astrophysical constraint~\cite{Candon:2024eah, Buschmann:2021juv} remains more stringent and already excludes the parameter space probed in this work.

The $^{23}$Na limits are the strongest but also the most dependent on the stellar model: they hold for Betelgeuse at the end of core carbon burning, and weaken by several orders of magnitude for earlier stages (see Fig.~\ref{fig:magnetic_stellar_uncertainty}). Unlike the magnetic conversion channel, which carries an additional uncertainty from the assumed Galactic magnetic field configuration, the decay channel is relatively more robust, as its limits depend mainly on the stellar model.

More broadly, this work demonstrates that a single nearby evolved star can serve as a source for several independent ALP searches spanning many decades in mass, using only archival X-ray and soft gamma-ray data. The method is not specific to Betelgeuse or to these isotopes, and deeper exposures, additional nuclear transitions, or next-generation MeV astronomy instruments such as COSI~\cite{Tomsick:2023aue} or newASTROGAM~\cite{Berge:2025kff} would extend its reach significantly in both channels.
\section*{Acknowledgments}
This work was supported by the Arthur B. McDonald Canadian Astroparticle Physics Research Institute, NSERC, the Canada Foundation for Innovation and the Ontario Ministry of Colleges, Universities, Research Excellence and Security. Computations were performed at the Queen's Centre for Advanced Computing. Research at Perimeter Institute is supported by the Government of Canada through the Department of Innovation, Science, and Economic Development, and by the Province of Ontario.
\newpage
\bibliography{apssamp}

@PREAMBLE{
 "\providecommand{\noopsort}[1]{}" 
 # "\providecommand{\singleletter}[1]{#1}%" 
}

@article{Carenza:2024ehj,
    author = "Carenza, Pierluca and Giannotti, Maurizio and Isern, Jordi and Mirizzi, Alessandro and Straniero, Oscar",
    title = "{Axion astrophysics}",
    eprint = "2411.02492",
    archivePrefix = "arXiv",
    primaryClass = "hep-ph",
    reportNumber = "BARI-TH/66-24",
    doi = "10.1016/j.physrep.2025.02.002",
    journal = "Phys. Rept.",
    volume = "1117",
    pages = "1--102",
    year = "2025"
}

@article{Raffelt:2006cw,
    author = "Raffelt, Georg G.",
    editor = "Kuster, Markus and Raffelt, Georg and Beltran, Berta",
    title = "{Astrophysical axion bounds}",
    eprint = "hep-ph/0611350",
    archivePrefix = "arXiv",
    reportNumber = "MPP-2006-172",
    doi = "10.1007/978-3-540-73518-2_3",
    journal = "Lect. Notes Phys.",
    volume = "741",
    pages = "51--71",
    year = "2008"
}

@article{Xiao_Carenza_Giannotti_Mirizzi_Perez_Straniero_Grefenstette,
title = {Betelgeuse constraints on coupling between axionlike particles and electrons},
author = {Xiao, M. and Carenza, P. and Giannotti, M. and Mirizzi, A. and Perez, K. M. and Straniero, O. and Grefenstette, B. W.},
journal = {Physical Review D},
volume = {106},
number = {12},
pages = {123019},
year = {2022},
url = {https://journals.aps.org/prd/abstract/10.1103/PhysRevD.106.123019},
doi = {10.1103/PhysRevD.106.123019}
}

@article{Marsh:2015xka,
    author = "Marsh, David J. E.",
    title = "{Axion Cosmology}",
    eprint = "1510.07633",
    archivePrefix = "arXiv",
    primaryClass = "astro-ph.CO",
    reportNumber = "KCL-PH-TH-2015-50",
    doi = "10.1016/j.physrep.2016.06.005",
    journal = "Phys. Rept.",
    volume = "643",
    pages = "1--79",
    year = "2016"
}

@book{Raffelt:1996wa,
    author = "Raffelt, G. G.",
    title = "{Stars as laboratories for fundamental physics}: {The astrophysics of neutrinos, axions, and other weakly interacting particles}",
    isbn = "978-0-226-70272-8",
    month = "5",
    year = "1996"
}

@article{Raffelt:1987im,
    author = "Raffelt, Georg and Stodolsky, Leo",
    title = "{Mixing of the Photon with Low Mass Particles}",
    reportNumber = "MPI-PAE/PTh-54/87",
    doi = "10.1103/PhysRevD.37.1237",
    journal = "Phys. Rev. D",
    volume = "37",
    pages = "1237",
    year = "1988"
}

@article{CUORE:2012ymr,
    author = "Alessandria, F. and others",
    collaboration = "CUORE",
    title = "{Search for 14.4 keV solar axions from M1 transition of Fe-57 with CUORE crystals}",
    eprint = "1209.2800",
    archivePrefix = "arXiv",
    primaryClass = "hep-ex",
    doi = "10.1088/1475-7516/2013/05/007",
    journal = "JCAP",
    volume = "05",
    pages = "007",
    year = "2013"
}

@article{DiLuzio:2021qct,
    author = "Di Luzio, Luca and others",
    title = "{Probing the axion{\textendash}nucleon coupling with the next generation of~axion helioscopes}",
    eprint = "2111.06407",
    archivePrefix = "arXiv",
    primaryClass = "hep-ph",
    reportNumber = "DESY-21-194",
    doi = "10.1140/epjc/s10052-022-10061-1",
    journal = "Eur. Phys. J. C",
    volume = "82",
    number = "2",
    pages = "120",
    year = "2022"
}

@article{Krcmar:1998xn,
    author = "Krcmar, M. and Krecak, Z. and Stipcevic, M. and Ljubicic, A. and Bradley, D. A.",
    title = "{Search for invisible axions using Fe-57}",
    eprint = "nucl-ex/9801005",
    archivePrefix = "arXiv",
    reportNumber = "IRB-ZEF-98-01",
    doi = "10.1016/S0370-2693(98)01231-3",
    journal = "Phys. Lett. B",
    volume = "442",
    pages = "38",
    year = "1998"
}

@article{Namba:2007rm,
    author = "Namba, T.",
    title = "{Results of a search for monochromatic solar axions using Fe-57}",
    doi = "10.1016/j.physletb.2007.01.005",
    journal = "Phys. Lett. B",
    volume = "645",
    pages = "398--401",
    year = "2007"
}

@article{Derbin:2009yb,
    author = "Derbin, A. V. and Egorov, A. I. and Mitropol'sky, I. A. and Muratova, V. N. and Semenov, D. A. and Unzhakov, E. V.",
    title = "{Search for Resonant Absorption of Solar Axions Emitted in M1 Transition in Fe-57 Nuclei}",
    eprint = "0906.0256",
    archivePrefix = "arXiv",
    primaryClass = "hep-ex",
    doi = "10.1140/epjc/s10052-009-1095-y",
    journal = "Eur. Phys. J. C",
    volume = "62",
    pages = "755--760",
    year = "2009"
}

@article{Jaeckel:2010ni,
    author = "Jaeckel, Joerg and Ringwald, Andreas",
    title = "{The Low-Energy Frontier of Particle Physics}",
    journal = "Ann. Rev. Nucl. Part. Sci.",
    volume = "60",
    pages = "405--437",
    year = "2010",
    doi = "10.1146/annurev.nucl.012809.104433",
    eprint = "1002.0329",
    archivePrefix = "arXiv",
    primaryClass = "hep-ph"
}

@article{Gaillard_Grannis_Sciulli,
  title = {The Standard Model of Particle Physics},
  author = {Gaillard, Mary K. and Grannis, Paul D. and Sciulli, Frank J.},
  journal = {Rev. Mod. Phys.},
  volume = {71},
  issue = {2},
  pages = {S96--S111},
  numpages = {16},
  year = {1999},
  month = {Mar},
  publisher = {American Physical Society},
  doi = {10.1103/RevModPhys.71.S96},
  url = {https://link.aps.org/doi/10.1103/RevModPhys.71.S96}
}

@article{Pecci_Quinn,
  title = {$\mathrm{CP}$ Conservation in the Presence of Pseudoparticles},
  author = {Peccei, R. D. and Quinn, Helen R.},
  journal = {Phys. Rev. Lett.},
  volume = {38},
  issue = {25},
  pages = {1440--1443},
  numpages = {0},
  year = {1977},
  month = {Jun},
  publisher = {American Physical Society},
  doi = {10.1103/PhysRevLett.38.1440},
  url = {https://link.aps.org/doi/10.1103/PhysRevLett.38.1440}
}

@article{Andriamonje2009CAST,
   title={Search for 14.4 keV solar axions emitted in the M1-transition of $^{57}$Fe nuclei with CAST},
   author={Andriamonje, S. and Arik, M. and Aune, S. and Autiero, D. and Barth, K. and Belov, A. and Bräuninger, H. and Cebrián, S. and Dafni, T. and Davenport, M. and Derbin, A. and Desch, K. and Di Lella, L. and Eleftheriadis, C. and Fanourakis, G. and Farach, H. and Ferretti, J. and Fischer, H. and Friedel, M. and Galán, J. and Garza, J. G. and Gehring, M. and Geralis, T. and Giomataris, I. and Gninenko, S. and Gómez, A. and Gómez-Cadenas, J. J. and Hasinoff, M. and Hoffmann, D. H. H. and Iguaz, F. and Irastorza, I. G. and Jacoby, J. and Jakovčić, K. and Kamyshkov, Y. and Kuster, M. and Lakić, B. and Lasseur, C. and Liolios, A. and Ljungberg, M. and Lounis, A. and Luzón, G. and Miller, E. H. and Morales, A. and Natale, G. and Neff, M. and Nikolaidou, R. and Ntanos, C. and Ortega, J. A. and Papaevangelou, T. and Pivovaroff, M. J. and Pugnat, P. and Ruz, J. and Ruzic, B. P. and Sakellariadou, M. and Semertzidis, Y. K. and Shilon, I. and Solovov, V. and Stewart, L. and Tomás, A. and Tsagri, M. and Tsuda, M. and Türler, M. and Vogel, J. K. and Vafeiadis, T. and Vavilov, S. and Viaux, N. and von Feilitzsch, F. and Walckiers, L. and Weber, M. and Yildiz, S. and Ziołkowski, M.},
   year={2009},
   month=sep,
   eprint={0906.4488},
   archivePrefix={arXiv},
   primaryClass={hep-ex},
   url={https://arxiv.org/abs/0906.4488},
   journal={Journal of Cosmology and Astroparticle Physics},
   volume={2009},
   number={12},
   pages={002}
}

@article{Cand_n_2025,
   title={Probing the axion-nucleon coupling with supergiant stars},
   volume={112},
   ISSN={2470-0029},
   url={http://dx.doi.org/10.1103/xv5c-j2rv},
   DOI={10.1103/xv5c-j2rv},
   number={12},
   journal={Physical Review D},
   publisher={American Physical Society (APS)},
   author={Candón, Francisco R. and Casaseca, Pablo and Giannotti, Maurizio and Kaltschmidt, Mathieu and Ruz, Jaime and Vogel, Julia K.},
   year={2025},
   month=Dec }

@article{Fleury:2022plh,
    author = "Fleury, Leesa and Caiazzo, Ilaria and Heyl, Jeremy",
    title = "{Constraining axions with ZTF J1901+1458}",
    eprint = "2208.00405",
    archivePrefix = "arXiv",
    primaryClass = "astro-ph.HE",
    doi = "10.1103/PhysRevD.107.L101303",
    journal = "Phys. Rev. D",
    volume = "107",
    number = "10",
    pages = "L101303",
    year = "2023"
}

@article{Ning:2025kyu,
    author = "Ning, Orion and Ray, Anupam and Safdi, Benjamin R.",
    title = "{Axion lines from nuclear de-excitations in galactic stellar populations}",
    eprint = "2509.03569",
    archivePrefix = "arXiv",
    primaryClass = "hep-ph",
    reportNumber = "CERN-TH-2025-162, N3AS-25-014",
    doi = "10.1103/xyjw-9bw7",
    journal = "Phys. Rev. D",
    volume = "113",
    number = "3",
    pages = "035010",
    year = "2026"
}

@article{Haxton:1991pu,
    author = "Haxton, W. C. and Lee, K. Y.",
    title = "{Red giant evolution, metallicity and new bounds on hadronic axions}",
    reportNumber = "DOE-ER-40427-2-N91",
    doi = "10.1103/PhysRevLett.66.2557",
    journal = "Phys. Rev. Lett.",
    volume = "66",
    pages = "2557--2560",
    year = "1991"
}

@misc{Nudat3,
  author = {{N. N. D. C. Center, Brookhaven National Laboratory}},
  title = {NuDat 3.0 Database},
  howpublished = {\url{https://www.nndc.bnl.gov/nudat3/}},
  year = {n.d.},
  note = {Brookhaven National Laboratory}
}

@software{2014ascl.soft08004N,
       author = {{Nasa High Energy Astrophysics Science Archive Research Center (Heasarc)}},
        title = "{HEAsoft: Unified Release of FTOOLS and XANADU}",
 howpublished = {Astrophysics Source Code Library, record ascl:1408.004},
         year = 2014,
        month = aug,
          eid = {ascl:1408.004},
archivePrefix = {ascl},
       eprint = {1408.004},
       adsurl = {https://ui.adsabs.harvard.edu/abs/2014ascl.soft08004N}
}

@article{Candon:2024eah,
    author = "Cand{\'o}n, Francisco R. and Fiorillo, Damiano F. G. and Lucente, Giuseppe and Vitagliano, Edoardo and Vogel, Julia K.",
    title = "{NuSTAR Bounds on Radiatively Decaying Particles from M82}",
    eprint = "2412.03660",
    archivePrefix = "arXiv",
    primaryClass = "hep-ph",
    doi = "10.1103/PhysRevLett.134.171004",
    journal = "Phys. Rev. Lett.",
    volume = "134",
    number = "17",
    pages = "171004",
    year = "2025"
}

@ARTICLE{2003A&A...411L...1W,
       author = {{Winkler}, C. and {Courvoisier}, T.~J.-L. and {Di Cocco}, G. and {Gehrels}, N. and {Gim{\'e}nez}, A. and {Grebenev}, S. and {Hermsen}, W. and {Mas-Hesse}, J.~M. and {Lebrun}, F. and {Lund}, N. and {Palumbo}, G.~G.~C. and {Paul}, J. and {Roques}, J.-P. and {Schnopper}, H. and {Sch{\"o}nfelder}, V. and {Sunyaev}, R. and {Teegarden}, B. and {Ubertini}, P. and {Vedrenne}, G. and {Dean}, A.~J.},
        title = "{The INTEGRAL mission}",
      journal = {\aap},
         year = 2003,
        month = nov,
       volume = {411},
        pages = {L1-L6},
          doi = {10.1051/0004-6361:20031288},
       adsurl = {https://ui.adsabs.harvard.edu/abs/2003A&A...411L...1W}
}

@article{MESA1,
   title={MODULES FOR EXPERIMENTS IN STELLAR ASTROPHYSICS (MESA)},
   volume={192},
   ISSN={1538-4365},
   url={http://dx.doi.org/10.1088/0067-0049/192/1/3},
   DOI={10.1088/0067-0049/192/1/3},
   number={1},
   journal={The Astrophysical Journal Supplement Series},
   publisher={American Astronomical Society},
   author={Paxton, Bill and Bildsten, Lars and Dotter, Aaron and Herwig, Falk and Lesaffre, Pierre and Timmes, Frank},
   year={2010},
   month=Dec, pages={3} }

@article{HarveySmith2011,
   title={Magnetic Fields in Large-Diameter H II Regions Revealed by the Faraday Rotation of Compact Extragalactic Radio Sources},
   volume={736},
   ISSN={0004-637X, 1538-4357},
   url={https://doi.org/10.1088/0004-637X/736/2/83},
   DOI={10.1088/0004-637X/736/2/83},
   number={2},
   journal={The Astrophysical Journal},
   publisher={American Astronomical Society},
   author={Harvey-Smith, L. and Madsen, G. J. and Gaensler, B. M.},
   year={2011},
   month=jul,
   pages={83}
}

@article{Moriyama,
title = {A Proposal to Search for a Monochromatic Component of Solar Axions Using 57Fe},
author = {Moriyama, Shigetaka},
journal = {Physical Review Letters},
year = {1995},
url = {https://arxiv.org/abs/hep-ph/9504318},
doi = {10.1103/PhysRevLett.75.3222}
}

@article{Haxton_Liu_McCutcheon_Ray,
title = {A Continuous Galactic Line Source of Axions: The Remarkable Case of 23Na},
author = {Haxton, W. C. and Liu, Xing and McCutcheon, Annie and Ray, Anupam},
journal = {arXiv:2505.03038 [astro-ph.HE]},
year = {2025},
url = {https://arxiv.org/abs/2505.03038},
doi = {10.48550/arXiv.2505.03038}
}

@article{Molnar_Joyce_Leung,
title = {Comment on the feasibility of carbon burning in Betelgeuse: a response to “The evolutionary stage of Betelgeuse inferred from its pulsation periods”},
author = {Molnár, László and Joyce, Meridith and Leung, Shing-Chi},
journal = {arXiv:2306.05600 [astro-ph.SR]},
year = {2023},
url = {https://arxiv.org/abs/2306.05600},
doi = {10.48550/arXiv.2306.05600}
}

@article{Haxton:2026ura,
    author = "Haxton, W. C. and Liu, Xing and Ray, Anupam and Rule, Evan",
    title = "{Is the Turner Window Open? Seeking Closure with Resonant Absorption of Galactic Axions in NaI Dark Matter Detectors}",
    eprint = "2603.16998",
    archivePrefix = "arXiv",
    primaryClass = "hep-ph",
    reportNumber = "N3AS-26-005, INT-26-008, LA-UR-26-22066",
    month = "3",
    year = "2026"
}

@article{Haxton:2026pty,
    author = "Haxton, W. C. and Liu, Xing and Ray, Anupam and Rule, Evan",
    title = "{Revisiting Turner Window Axions: The Untapped Potential of NaI Dark Matter Detectors}",
    eprint = "2604.26004",
    archivePrefix = "arXiv",
    primaryClass = "hep-ph",
    reportNumber = "N3AS-26-007, INT-PUB-26-016, LA-UR-26-21773",
    month = "4",
    year = "2026"
}

@article{Kumar:2026nlg,
    author = "Kumar, Tanmoy and Nath, Newton",
    title = "{Solar Axions from Nuclear Transitions}",
    eprint = "2605.24121",
    archivePrefix = "arXiv",
    primaryClass = "hep-ph",
    month = "5",
    year = "2026"
}

@article{Saio_Nandal_Meynet_Ekstom,
title = {The evolutionary stage of Betelgeuse inferred from its pulsation periods},
author = {Saio, Hideyuki and Nandal, Devesh and Meynet, Georges and Ekström, Sylvia},
journal = {arXiv:2306.00287 [astro-ph.SR]},
year = {2023},
url = {https://arxiv.org/abs/2306.00287},
doi = {10.48550/arXiv.2306.00287}
}

@article{ParticleDataGroup:2022pth,
    author = "Workman, R. L. and others",
    collaboration = "Particle Data Group",
    title = "{Review of Particle Physics}",
    doi = "10.1093/ptep/ptac097",
    journal = "PTEP",
    volume = "2022",
    pages = "083C01",
    year = "2022"
}

@article{Berge:2025kff,
    author = "Berge, David and Mazziotta, Mario Nicola and Tavani, Marco and Tatischeff, Vincent and Oberlack, Uwe",
    title = "{newASTROGAM: The New MeV to GeV Gamma-ray Observatory}",
    eprint = "2507.08133",
    archivePrefix = "arXiv",
    primaryClass = "astro-ph.IM",
    doi = "10.22323/1.501.0572",
    journal = "PoS",
    volume = "ICRC2025",
    pages = "572",
    year = "2025"
}

@article{Tomsick:2023aue,
    author = "Tomsick, John A. and others",
    title = "{The Compton Spectrometer and Imager}",
    eprint = "2308.12362",
    archivePrefix = "arXiv",
    primaryClass = "astro-ph.HE",
    doi = "10.22323/1.444.0745",
    journal = "PoS",
    volume = "ICRC2023",
    pages = "745",
    year = "2023"
}

@article{Fiorillo:2026rgo,
    author = "Fiorillo, Damiano F. G. and Gil Muyor, {\'A}ngel and Raffelt, Georg G. and Vitagliano, Edoardo",
    title = "{Magnetic Turbulence Boosts Supernova Signals of Axion-Photon Conversion}",
    eprint = "2605.15261",
    archivePrefix = "arXiv",
    primaryClass = "hep-ph",
    doi = "10.1103/wdbg-66y1",
    journal = "Phys. Rev. Lett.",
    volume = "137",
    number = "12",
    pages = "121002",
    year = "2026"
}

@article{NuSTAR:2013yza,
    author = "Harrison, Fiona A. and others",
    collaboration = "NuSTAR",
    title = "{The Nuclear Spectroscopic Telescope Array (NuSTAR) High-Energy X-Ray Mission}",
    eprint = "1301.7307",
    archivePrefix = "arXiv",
    primaryClass = "astro-ph.IM",
    reportNumber = "SLAC-PUB-16148",
    doi = "10.1088/0004-637X/770/2/103",
    journal = "Astrophys. J.",
    volume = "770",
    pages = "103",
    year = "2013"
}

@article{Madsen:2015jea,
    author = "Madsen, Kristin K. and others",
    title = "{Calibration of the NuSTAR High Energy Focusing X-ray Telescope}",
    eprint = "1504.01672",
    archivePrefix = "arXiv",
    primaryClass = "astro-ph.IM",
    doi = "10.1088/0067-0049/220/1/8",
    journal = "Astrophys. J. Suppl.",
    volume = "220",
    number = "1",
    pages = "8",
    year = "2015"
}

@article{Unger:2023lob,
    author = "Unger, Michael and Farrar, Glennys R.",
    title = "{The Coherent Magnetic Field of the Milky Way}",
    eprint = "2311.12120",
    archivePrefix = "arXiv",
    primaryClass = "astro-ph.GA",
    doi = "10.3847/1538-4357/ad4a54",
    journal = "Astrophys. J.",
    volume = "970",
    number = "1",
    pages = "95",
    year = "2024"
}

@article{ParticleDataGroup:2024cfk,
    author = "Navas, S. and others",
    collaboration = "Particle Data Group",
    title = "{Review of particle physics}",
    doi = "10.1103/PhysRevD.110.030001",
    journal = "Phys. Rev. D",
    volume = "110",
    number = "3",
    pages = "030001",
    year = "2024"
}

@article{Ning:2024eky,
    author = "Ning, Orion and Safdi, Benjamin R.",
    title = "{Leading Axion-Photon Sensitivity with NuSTAR Observations of M82 and M87}",
    eprint = "2404.14476",
    archivePrefix = "arXiv",
    primaryClass = "hep-ph",
    doi = "10.1103/PhysRevLett.134.171003",
    journal = "Phys. Rev. Lett.",
    volume = "134",
    number = "17",
    pages = "171003",
    year = "2025"
}

@article{Buschmann:2021juv,
    author = "Buschmann, Malte and Dessert, Christopher and Foster, Joshua W. and Long, Andrew J. and Safdi, Benjamin R.",
    title = "{Upper Limit on the QCD Axion Mass from Isolated Neutron Star Cooling}",
    eprint = "2111.09892",
    archivePrefix = "arXiv",
    primaryClass = "hep-ph",
    doi = "10.1103/PhysRevLett.128.091102",
    journal = "Phys. Rev. Lett.",
    volume = "128",
    number = "9",
    pages = "091102",
    year = "2022"
}

@article{Cowan:2011an,
    author = "Cowan, Glen and Cranmer, Kyle and Gross, Eilam and Vitells, Ofer",
    title = "{Power-Constrained Limits}",
    eprint = "1105.3166",
    archivePrefix = "arXiv",
    primaryClass = "physics.data-an",
    month = "5",
    year = "2011"
}

@article{Kuulkers:2021svd,
    author = "Kuulkers, Erik and others",
    title = "{INTEGRAL reloaded: Spacecraft, instruments and ground system}",
    eprint = "2106.12446",
    archivePrefix = "arXiv",
    primaryClass = "astro-ph.IM",
    doi = "10.1016/j.newar.2021.101629",
    journal = "New Astron. Rev.",
    volume = "93",
    pages = "101629",
    year = "2021"
}

@article{Courvoisier:2003vm,
    author = "Courvoisier, T. J. L. and others",
    title = "{The INTEGRAL Science Data Centre (ISDC)}",
    eprint = "astro-ph/0308047",
    archivePrefix = "arXiv",
    doi = "10.1051/0004-6361:20031172",
    journal = "Astron. Astrophys.",
    volume = "411",
    pages = "L53--L57",
    year = "2003"
}

@article{Dolan_2016,
   title={EVOLUTIONARY TRACKS FOR BETELGEUSE},
   volume={819},
   ISSN={1538-4357},
   url={http://dx.doi.org/10.3847/0004-637X/819/1/7},
   DOI={10.3847/0004-637x/819/1/7},
   number={1},
   journal={The Astrophysical Journal},
   publisher={American Astronomical Society},
   author={Dolan, Michelle M. and Mathews, Grant J. and Lam, Doan Duc and Lan, Nguyen Quynh and Herczeg, Gregory J. and Dearborn, David S. P.},
   year={2016},
   month=Feb, pages={7} }

@article{Goldberg:2024dml,
    author = "Goldberg, Jared A. and Joyce, Meridith and Moln{\'a}r, L{\'a}szl{\'o}",
    title = "{A Buddy for Betelgeuse: Binarity as the Origin of the Long Secondary Period in {\ensuremath{\alpha}} Orionis}",
    eprint = "2408.09089",
    archivePrefix = "arXiv",
    primaryClass = "astro-ph.SR",
    doi = "10.3847/1538-4357/ad87f4",
    journal = "Astrophys. J.",
    volume = "977",
    number = "1",
    pages = "35",
    year = "2024"
}

@article{Luo_2022,
   title={Stellar Models of Betelgeuse Constrained Using Observed Surface Conditions},
   volume={927},
   ISSN={1538-4357},
   url={http://dx.doi.org/10.3847/1538-4357/ac4f5f},
   DOI={10.3847/1538-4357/ac4f5f},
   number={1},
   journal={The Astrophysical Journal},
   publisher={American Astronomical Society},
   author={Luo, Tianyin and Umeda, Hideyuki and Yoshida, Takashi and Takahashi, Koh},
   year={2022},
   month=Mar, pages={115} }

@article{Cowan:2010js,
    author = "Cowan, Glen and Cranmer, Kyle and Gross, Eilam and Vitells, Ofer",
    title = "{Asymptotic formulae for likelihood-based tests of new physics}",
    eprint = "1007.1727",
    archivePrefix = "arXiv",
    primaryClass = "physics.data-an",
    doi = "10.1140/epjc/s10052-011-1554-0",
    journal = "Eur. Phys. J. C",
    volume = "71",
    pages = "1554",
    year = "2011",
    note = "[Erratum: Eur.Phys.J.C 73, 2501 (2013)]"
}

@article{Fiorillo:2025gnd,
    author = "Fiorillo, Damiano F. G. and Gil Muyor, {\'A}ngel and Janka, Hans-Thomas and Raffelt, Georg G. and Vitagliano, Edoardo",
    title = "{Axion-photon conversion in transient compact stars: Systematics, constraints, and opportunities}",
    eprint = "2509.13322",
    archivePrefix = "arXiv",
    primaryClass = "hep-ph",
    doi = "10.1088/1475-7516/2026/03/053",
    journal = "JCAP",
    volume = "03",
    pages = "053",
    year = "2026"
}
\appendix
\section{Magnetic Field Uncertainty}
\label{bfield_uncertainty}
\begin{figure*}[t]
    \centering

\begin{minipage}[t]{0.49\textwidth}
        \centering
        \includegraphics[width=\linewidth]{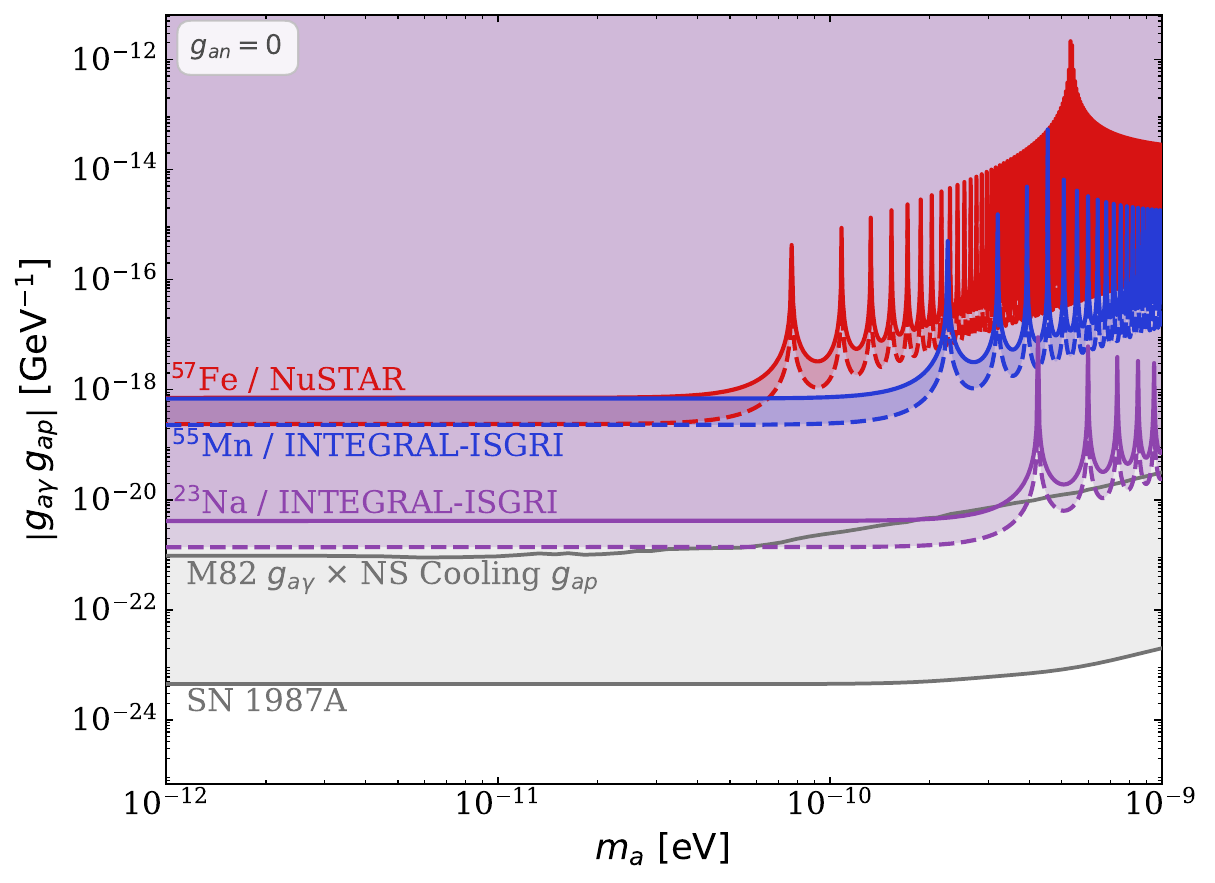}
        \label{fig:primakoff_shaded_proton}
\end{minipage}
\hfill
\begin{minipage}[t]{0.49\textwidth}
      \centering
        \includegraphics[width=\linewidth]{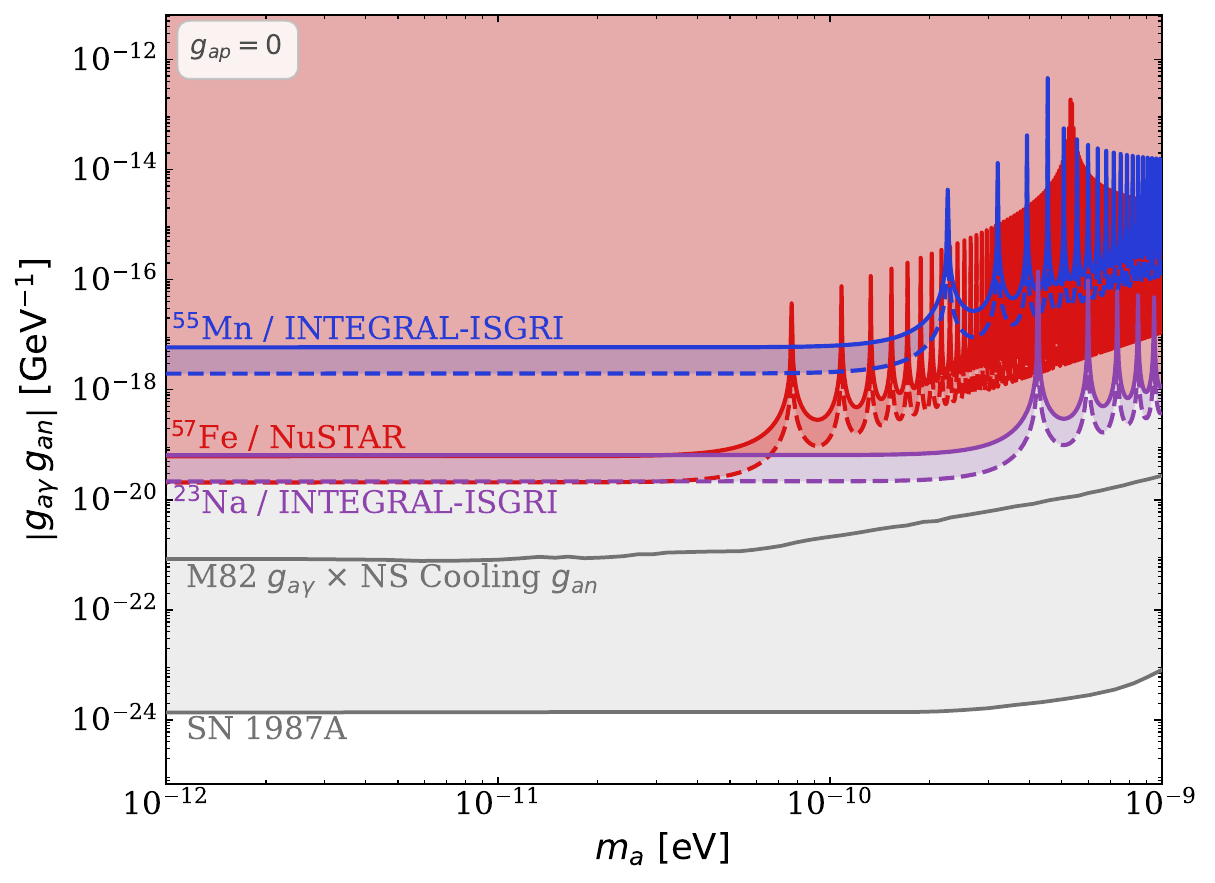}
        \label{fig:primakoff_shaded_neutron}
\end{minipage}
    
\caption{95\% CL exclusion limits on $|g_{a\gamma}g_{ap}|$ (left, $g_{an}=0$) and $|g_{a\gamma}g_{an}|$ (right, $g_{ap}=0$) for $^{23}$Na/INTEGRAL-ISGRI (purple), $^{55}$Mn/INTEGRAL-ISGRI (blue) and $^{57}$Fe/NuSTAR (red), for the fiducial Betelgeuse model at the end of core carbon burning ($T_c=95$~keV). Solid lines correspond to a Galactic magnetic field $B_T=1\,\mu$G and dashed lines to the more optimistic $B_T=3\,\mu$G; the band between them shows the magnetic-field uncertainty. The coloured areas are excluded by our analysis. The grey curves are existing bounds on the same coupling products~\cite{Fiorillo:2025gnd,Fiorillo:2026rgo, Ning:2025kyu, Ning:2024eky,Buschmann:2021juv}. See text for more details.}
    \label{fig:primakoff_shaded}
\end{figure*}

The $a\to\gamma$ magnetic conversion constraints derived in this work depend on the Galactic magnetic field transverse to the line of sight to Betelgeuse ($B_T$), whose strength carries a considerable uncertainty. We adopt a fiducial conservative value of $B_{T} = 1.0\,\mu\mathrm{G}$~\cite{Unger:2023lob}, although values up to $B_{T} = 3.0\,\mu\mathrm{G}$ are plausible~\cite{HarveySmith2011}. Since the ALP-photon conversion probability scales as $P_{a\to\gamma}\propto B_T^2$, the limits on $|g_{a\gamma}g_{aN}^{\rm eff}|$ scale as $1/B_T$. The optimistic value of $B_T$ therefore strengthens the
  limits by a factor of three at all masses. This dependence is illustrated in Fig.~\ref{fig:primakoff_shaded} (see also Fig.~\ref{fig:primakoff_final}), where the solid boundary corresponds to $B_{T} = 1.0\,\mu\mathrm{G}$ (conservative), the dashed boundary to $B_{T} = 3.0\,\mu\mathrm{G}$ (optimistic), and the shaded band spans the full range.
                              
\begin{figure*}
    \centering

    \begin{minipage}[t]{0.49\textwidth}
        \centering
        \includegraphics[width=\linewidth]{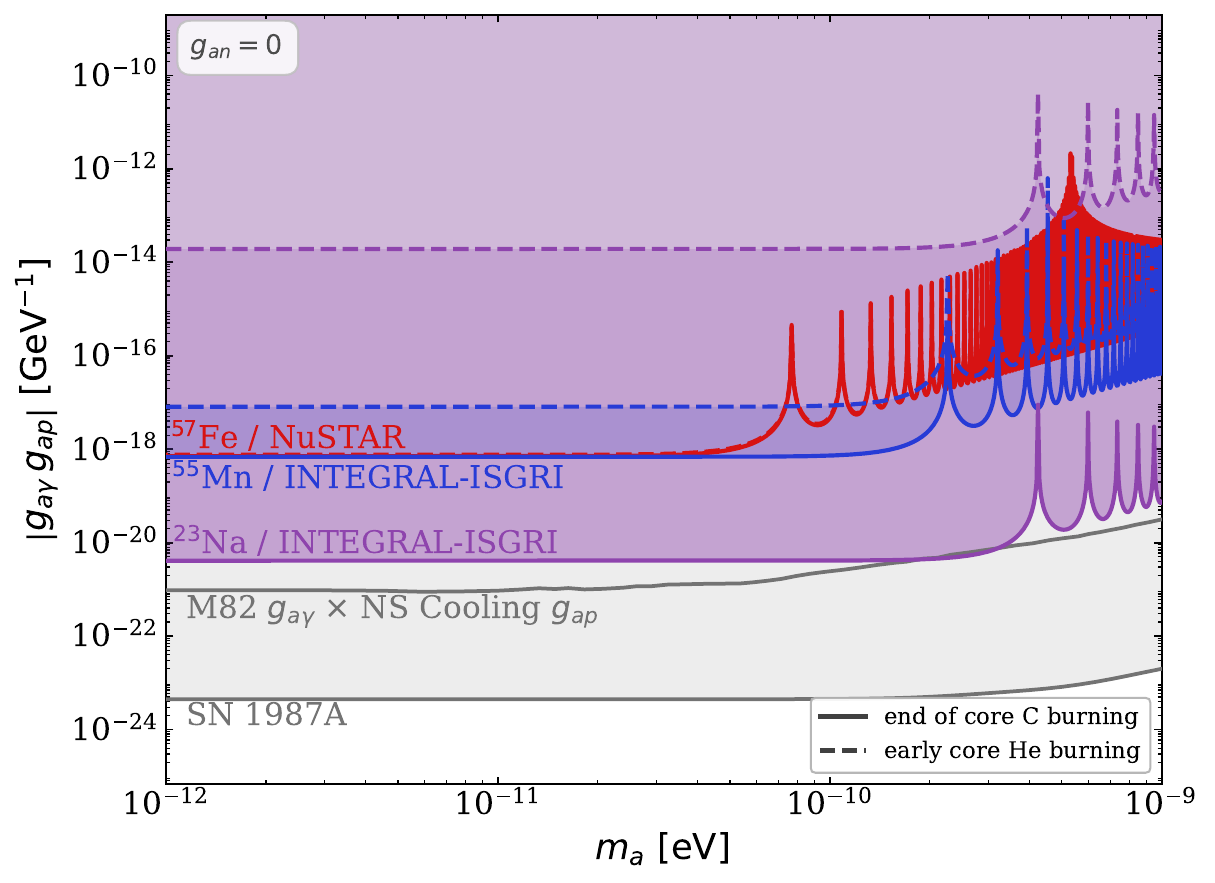}
        \label{fig:primakoff_stage_proton}
    \end{minipage}
    \hfill
    \begin{minipage}[t]{0.49\textwidth}
        \centering
        \includegraphics[width=\linewidth]{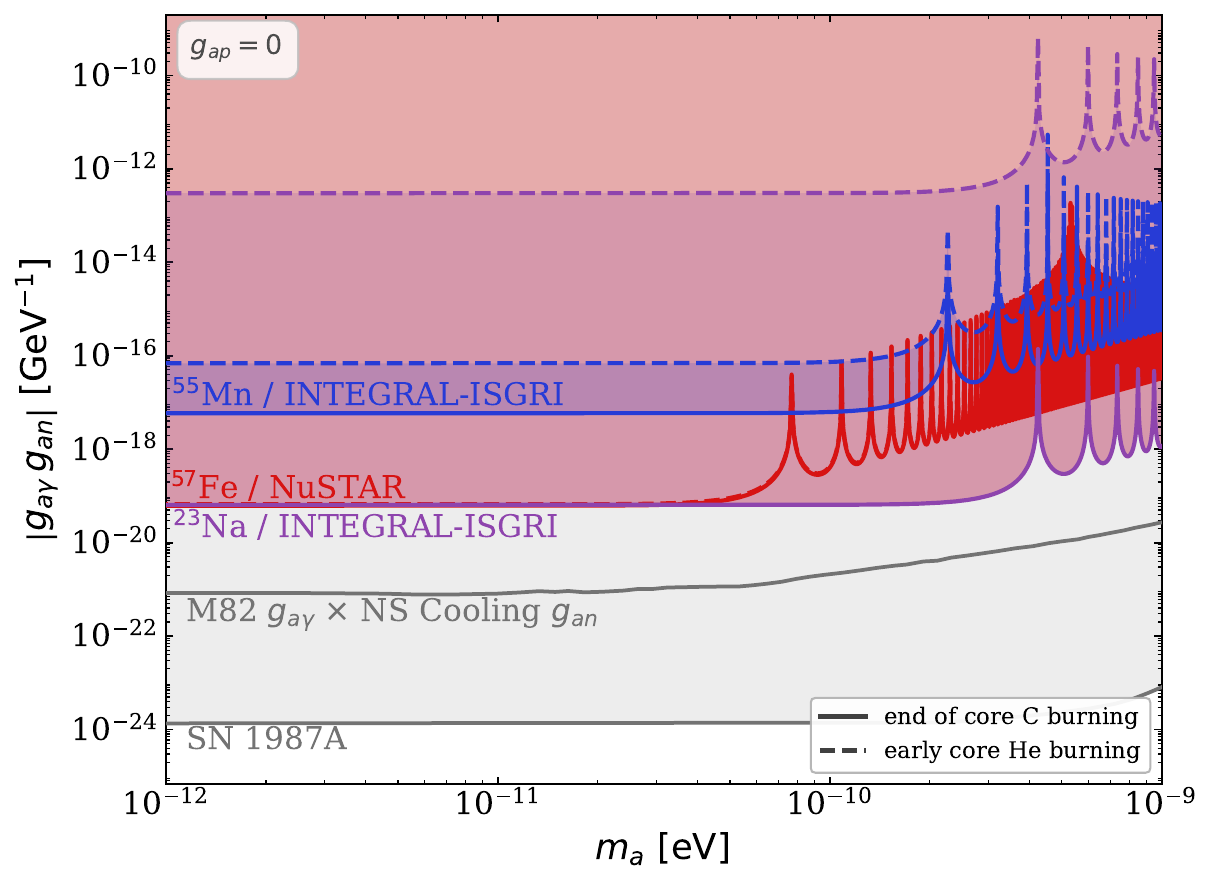}
        \label{fig:primakoff_stage_neutron}
    \end{minipage}

    \caption{Dependence of the magnetic-conversion limits on the evolutionary stage of Betelgeuse: 95\% CL exclusion limits on $|g_{a\gamma}g_{ap}|$ (left, $g_{an}=0$) and $|g_{a\gamma}g_{an}|$
      (right, $g_{ap}=0$) for $^{23}$Na/INTEGRAL-ISGRI (purple), $^{55}$Mn/INTEGRAL-ISGRI (blue) and $^{57}$Fe/NuSTAR (red), for a Galactic magnetic field $B_T=1\,\mu$G. Solid lines correspond to
      the fiducial model at the end of core carbon burning ($T_c=95$~keV) and dashed lines to early core helium burning ($T_c=15$~keV). The darker coloured areas are excluded at both stages, the lighter ones
  only at the end of core carbon burning. The grey curves are existing bounds on the same coupling products~\cite{Fiorillo:2025gnd,Fiorillo:2026rgo, Ning:2025kyu, Ning:2024eky,Buschmann:2021juv}. See text for more details.}
    \label{fig:magnetic_stellar_uncertainty}
\end{figure*}
\section{Evolutionary Stage Uncertainty}
\label{uncertainty}
\begin{figure}
    \centering
    \includegraphics[width=\columnwidth]{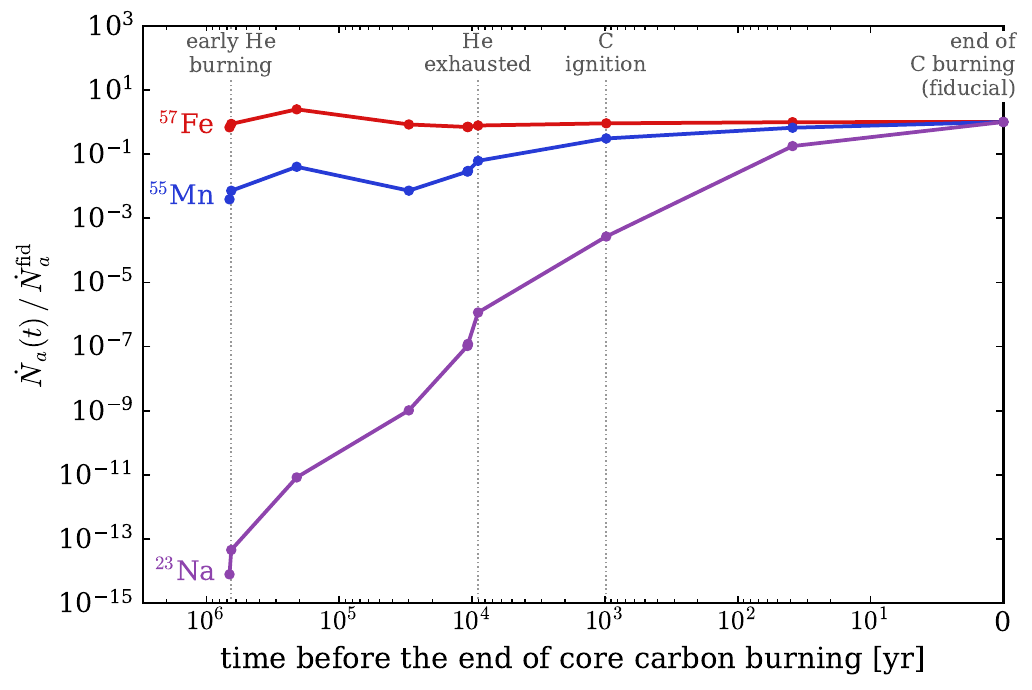}
    \caption{Total ALP emission rate of $^{57}$Fe (red), $^{55}$Mn (blue) and $^{23}$Na (purple) along
    the evolution of the $20\,M_\odot$, $Z=0.02$ MESA model, relative to the fiducial model at the end of
    core carbon burning.}
    \label{fig:emissivity_time}
  \end{figure}

The constraints on ALP couplings derived in this work (both magnetic field conversion and decay) depend on the assumed evolutionary stage of Betelgeuse, which sets the core temperature, density and isotopic abundances of the stellar model, and hence the thermal population of the nuclear excited states that source the ALP flux. The current evolutionary state of Betelgeuse remains debated: some studies place it in core helium burning~\cite{Molnar_Joyce_Leung, Dolan_2016, Goldberg:2024dml}, while others place it in or beyond core carbon burning~\cite{Saio_Nandal_Meynet_Ekstom, Luo_2022}. The stage is difficult to determine because the late burning phases are short compared with the thermal timescale of the envelope, so the surface barely responds to changes in the core. In our model, the luminosity changes by only 2\% and the effective temperature by less than 10~K between carbon ignition and the end of core carbon burning, while the central temperature rises from 56 to 95~keV. Throughout the analysis we adopt a fiducial MESA model of a $20\,M_\odot$, $Z=0.02$ star at the end of core carbon burning~\cite{Cand_n_2025}. The stage dependence of the limits is primarily controlled by the Boltzmann factor $e^{-E_*/T}$: isotopes with larger transition energies $E_*$ are exponentially more sensitive to the core temperature $T_c$. Between early core helium burning ($T_c\simeq15$~keV) and the end of core carbon burning ($T_c\simeq95$~keV), the Boltzmann factor alone changes by $\sim2\times10^{10}$ for $^{23}$Na ($E_*=440.2$~keV), $\sim10^{3}$ for $^{55}$Mn (125.9~keV), and only by a factor of $\simeq2$ for $^{57}$Fe (14.4~keV), whose level is already substantially populated during helium burning. Including the changes in density and abundance, the total ALP emission rate changes by factors of $2\times10^{13}$, $1.4\times10^{2}$ and $1.15$, respectively (Fig.~\ref{fig:emissivity_time}), so the magnetic-conversion limits on $|g_{a\gamma}g_{aN}^{\rm eff}|$ weaken by factors of $5\times10^{6}$, 12 and 1.07 (Fig.~\ref{fig:magnetic_stellar_uncertainty}, where the solid boundary corresponds to the end of core carbon burning and the dashed boundary to early core helium burning). Thus, although our $^{23}$Na-based limits are the strongest in both channels, they hold only if Betelgeuse is in, or close to, its carbon-burning phase, whereas the $^{55}$Mn- and, in particular, the $^{57}$Fe-based limits are considerably more robust, since their lower transition energies make the excited-state population much less sensitive to the core temperature.
\bibliographystyle{JHEP}
\end{document}